\documentclass[twocolumn]{aastex631}

\usepackage{mathptmx}
\usepackage[T1]{fontenc}
\usepackage{ae,aecompl}
\usepackage{graphicx}	
\usepackage{subfigure}
\usepackage{booktabs} 
\usepackage{subfloat}
\usepackage{amsmath}	
\usepackage{amssymb}	
\usepackage{gensymb}    
\usepackage{enumerate}
\usepackage{multirow}
\usepackage{natbib}

\usepackage{chngcntr}
\usepackage{ulem}
\usepackage{appendix}
\usepackage{CJK}
\usepackage{threeparttable}
\usepackage{xcolor}
\usepackage{makecell}

\shorttitle{Li and spots}
\shortauthors{Luo et al}

\begin{document}
\begin{CJK*}{UTF8}{gbsn}

\title{The Contribution of Starspots to the Lithium Spread in NGC~2264}

\author{Yan Luo (骆彦)}
\affiliation{Department of Astronomy, Peking University, Yiheyuan Lu 5, Haidian Qu, 100871 Beijing, People's Republic of China}
\affiliation{Kavli Institute for Astronomy and Astrophysics, Peking University, Yiheyuan Lu 5, Haidian Qu, 100871 Beijing, People's Republic of China}

\author[0000-0002-7154-6065]{Gregory J. Herczeg (沈雷歌）}
\affiliation{Department of Astronomy, Peking University, Yiheyuan Lu 5, Haidian Qu, 100871 Beijing, People's Republic of China}
\affiliation{Kavli Institute for Astronomy and Astrophysics, Peking University, Yiheyuan Lu 5, Haidian Qu, 100871 Beijing, People's Republic of China}

\author{Min Fang}
\affiliation{Purple Mountain Observatory, Chinese Academy of Sciences, 10 Yuanhua Road, Nanjing 210023, People's Republic of China}
\affiliation{University of Science and Technology of China, Hefei 230026, People's Republic of China}

\author{Qinghui Sun}
\affiliation{Tsung-Dao Lee Institute, Shanghai Jiao Tong University, Shanghai, 200240, People's Republic of China}

\author[0000-0002-6506-1985]{Xiaoting Fu (符晓婷)}
\affiliation{Purple Mountain Observatory, Chinese Academy of Sciences, 10 Yuanhua Road, Nanjing 210023, People's Republic of China}
\affiliation{University of Science and Technology of China, Hefei 230026, People's Republic of China}
\affiliation{Nanjing Key Laboratory for Space and Underground Astrophysics, Nanjing 210023, China}

\correspondingauthor{Gregory Herczeg, gherczeg1@gmail.com and Yan Luo, yanluo@stu.pku.edu.cn}


\begin{abstract}
Star clusters with ages of a few Myr have a previously unexplained dispersion in the lithium equivalent widths. In this work, we show that much of the lithium spread of the young cluster NGC~2264 can be attributed to the observational presence of starspots. We demonstrate that the equivalent width of \ion{Li}{1}\,6708\,\AA\ changes with observed spot coverage, using a sample of 15 Weak-Lined T~Tauri Stars (WTTSs) with multi-epoch high-resolution ESPaDOnS spectra and photometry. 
{To quantify the contribution of starspots, we construct an empirical two-component starspot model. When differences in visible spot coverage are taken into account, 68\% of the WTTSs fall within the model-predicted EW(Li) dispersion for a range from 0-90\% visible spot coverage.
We find no correlation between the age spread or gravity and \ion{Li}{1} equivalent width in NGC~2264.
Our results suggest that invoking starspots can explain much of the apparent lithium spread in NGC~2264 and reduces but does not fully eliminate the need to invoke rotation-driven lithium depletion or other non-standard explanations at such young ages.  Future simultaneous measurements of lithium and beryllium in young stars may provide an independent test of this scenario.}
\end{abstract}

\section{Introduction}
Lithium is a fragile element that burns at temperatures of {a few million K, which the bottom of the convective zone reaches during pre-main sequence contraction.} For low-mass stars, convective mixing cycles lithium to the core, leading to the rapid depletion of lithium through the entire star \citep{bildsten1997lithium,fu2015}. Because lithium is destroyed as the star contracts, the depletion of lithium serves as a sensitive diagnostic of the age of young clusters, providing constraints on pre-main sequence evolution and star formation histories (see review by \citealp{jeffries2014using}).

While models predict depletion at specific ages, measurements of lithium line strengths demonstrate that depletion depends on properties beyond just stellar mass and age.   
A dispersion in the lithium equivalent widths of the $\sim$120\,Myr old Pleiades cluster was originally identified by \citet{duncan1983lithium}. Fast rotators tend to be more lithium-rich than slow rotators, indicating a correlation between lithium and rotation  \citep{soderblom1993evolution}.
The lithium-rotation connection in the Pleiades has been attributed either to a real difference in Li abundance, potentially caused by 1) radius inflation \citep[e.g.][]{somers2014rotation,somers2015rotation,somers2017measurement}, 2) extra mixing processes \citep[e.g.][]{baraffe2017lithium}, 3) lasting differences in the interaction with protoplanetary disks (see e.g. \citealp{bouvier2018li} and references therein), or to the difference in line strengths due to magnetic activities \citep{leone2007magnetic} and surface inhomogeneities \citep{stuik1996modeling,king2000lithium, king2004alkali}.

\begin{table*}[!t]
    \centering
\caption{Photospheric properties of the variability sample}
    \begin{tabular}{lcccccccc}
    \toprule
    Source Name & SpT &${\rm T_{phot}}$ & Period & ${\rm Li_{EW},std}$  & ${\rm Vmag, std} $ & ${\rm TiO, std}$ & $JD_{spec}-\mathrm{2.45e6}$ & $JD_{phot}-\mathrm{2.45e6}$ \\
     Unit & & K & day & m\AA && & day& day\\
    \midrule
    LkCa 4 & M1.3\textsuperscript{[a]} & 4100$\pm$50\textsuperscript{[A]} & 3.374 & 634, 13 & 12.85, 0.12 & 1.478, 0.041 & 6665.7--6678.9 & 6594.0--6980.9 \\
    LkCa 7 & M1.2\textsuperscript{[a]} & &5.647 & 565, 12 & 12.31, 0.15 & 1.275, 0.076 & 7344.8--7359.8 & 7222.1--7602.1 \\
    Par 1379 & K4.0\textsuperscript{[b]} & 4600$\pm$50\textsuperscript{[B]}&5.676 & 471,  8 & 12.94, 0.05 & 1.063, 0.005 & 6665.9--6678.9 & 6594.0-7252.9 \\
    ROX 39 & K5.0\textsuperscript{[c]} & &0.882 & 569, 14 & 12.66, 0.04 & 1.167, 0.007 & 7548.8--7907.9 & 7252.5--7877.8 \\
    ROXs 45F & K5.0\textsuperscript{[d]} & &2.454 & 482,  9 & 12.68, 0.03 & 1.157, 0.013 & 7107.1--7123.1 & 6885.5--7278.5 \\
    RX J1608.0-3857 & M0.0\textsuperscript{[e]} & &2.425 & 636,  8 & 12.93, 0.05 & 1.319, 0.017 & 6428.9--6474.8 & 7456.8--7973.6 \\
    RX J1609.5-3850 & M0.5\textsuperscript{[f]} & &3.868 & 544, 14 & 12.64, 0.06 & 1.225, 0.043 & 6429.0--6443.9 & 7456.8--7975.6 \\
    TAP~26 & K4.0\textsuperscript{[a]} & 4620$\pm$50\textsuperscript{[C]}&0.714 & 411, 16 & 12.24, 0.04 & 1.090, 0.007 & 7344.9--9576.9 & 7366.8--8028.0 \\
    TAP~45 & K6.0\textsuperscript{[a]} & &9.587 & 511,  5 & 13.07, 0.04 & 1.183, 0.011 & 7345.0--7360.0 & 7274.1--7443.8 \\
    TWA 25 & M0.5\textsuperscript{[a]} & 4120$\pm$50\textsuperscript{[D]}&5.065 & 536, 13 & 11.29, 0.13 & 1.390, 0.055 & 7497.8--7887.8 & 7821.6--7944.5 \\
    TWA 6 & M0.0\textsuperscript{[a]} & 4425$\pm$50\textsuperscript{[E]}&0.541 & 471, 22 & 11.38, 0.10 & 1.181, 0.024 & 4169.9--6710.0 & 6790.6--7018.8 \\
    TWA 8a & M2.9\textsuperscript{[a]} & 3800$\pm$150\textsuperscript{[E]}&4.630 & 554,  6 & 12.33, 0.04 & 2.189, 0.031 & 7107.9--7122.9 & 6840.5--7390.9 \\
    V410 Tau & K3.0\textsuperscript{[g]} & 4500$\pm$100\textsuperscript{[F]}&1.872 & 500, 13 & 10.86, 0.03 & 1.131, 0.006 & 4755.0--4845.7 & 4682.5--4858.2 \\
    V819 Tau & K8.0\textsuperscript{[a]} & 4250$\pm$50\textsuperscript{[G]}&5.535 & 590,  9 & 13.05, 0.16 & 1.193, 0.027 & 5486.1--7037.8 & 6918.0--7072.8 \\
    V830 Tau & K7.5\textsuperscript{[a]} & 4250$\pm$50\textsuperscript{[G]}&2.744 & 607, 12 & 12.30, 0.07 & 1.198, 0.016 & 7011.9--7826.8 & 7102.7--7720.0 \\
    \bottomrule
    \end{tabular}

    \vspace{0.5ex}
    \begin{minipage}{\textwidth}
    \footnotesize
    \raggedright
    The spectral types are derived from:
    [a] \citet{herczeg2014optical};
    [b] \citet{hill2017magnetic};
    [c] \citet{wahhaj2010spitzer};
    [d] \citet{pecaut2016star};
    [e] \citet{galli2015evolution};
    [f] \citet{luhman2020gaia};
    [g] \citet{fernandez1998spectroscopy}.

    \smallskip
    The photosphere temperatures are derived from high-resolution optical spectra:
    [A] \citet{Donati2014MaTYSSE};
    [B] \citet{hill2017magnetic};
    [C] \citet{Yu2017tap26};
    [D] \citet{Nicholson2021twa25};
    [E] \citet{Hill2019twa};
    [F] \citet{yu2019v410};
    [G] \citet{Donati2015v819830}.
    \end{minipage}

    \label{tab:wtts}
\end{table*}

The dispersion in Li abundance has also been observed in younger associations, including $\alpha$ Persei ($\sim$50 Myr, \citealp{xiong2005lithium}), IC~2602 and IC~2391 (30--50 Myr, \citealp{randich2001ic}), NGC~2547 ($\sim$35 Myr, \citealp{binks2022ngc}),  $\beta$ Pic ($\sim$25 Myr, \citealp{messina2016rotation}), $\gamma$ Velorum cluster ($\sim$16 Myr, \citealp{Jeffries2017vel,jackson2025growth}), the Lupus star-forming region (1--7 Myr, \citealp{biazzo2017lup}), Collinder~69 ($\sim$5 Myr, \citealp{bayo2011coll}), NGC~2264 (6 Myr, \citealp{bouvier2016gaia}), and $\sigma$ Orionis (2--4 Myr, \citealp{zapatero2002ori}), with the lithium dispersion increasing with age between 10--50\,Myr \citep{jackson2025growth}.

The detection of the lithium dispersion and correlation with rotation in NGC~2264 \citep{bouvier2016gaia} suggests that the difference in Li abundance in older clusters may already be in place at young ages. However, NGC~2264 is thought to be younger than the lithium depletion timescale of 10--20 Myr \citep{jackson2014effect, somers2020spots, jackson2025growth}, so a real difference in abundance is challenging to explain. \citet{bouvier2016gaia} suggests that this early relation between lithium and rotation is most likely explained by radius inflation, though they do not rule out links with accretion history, planet formation, and angular momentum evolution. 

As an alternative to real differences in Li abundance, the observed spread in equivalent widths may instead be introduced by the visual appearance of starspots. Cool spots in young low-mass stars are evident in the periodic modulation of brightness caused by spot inhomogeneities {\citep[e.g.][]{bouvier1993rot,herbst1994catalogue,2005XiongSpot,grankin2008}}. {The \ion{Li}{1} line has a larger equivalent width in cooler regions where the ionization fraction is low, as seen in solar spots \citep{giampapa84}, despite spots and plage having indistinguishable Li abundances \citep{ritzenhoff97}.}
Spectroscopic analyses using two photospheric temperatures have revealed that cool spots can cover large fractions of the visible surface \citep[e.g.][]{gully2017placing,cao2022spot,perez2025spectral,perez2025igrins}. The change of spot fraction caused by stellar rotation should modulate lithium equivalent widths, since the line is stronger for cooler photospheres {\citep[e.g.][]{Pavlenko1996ewli,2005XiongSpot,Soriano2015li}}. This possibility is supported by the recent work of \citet{Franciosini2022pmsLi}, who analyzed five clusters (ages 10--100\,Myr) from the Gaia-ESO Survey and demonstrated that pre-main sequence models incorporating starspots can simultaneously reproduce the observed color-magnitude diagrams and lithium depletion patterns. Their findings suggest that a range of spot coverage, potentially varying with mass, is required to explain the lithium dispersion. 
{ 
Their work focused on the effect of starspots on stellar structure and Li burning through radius inflation. Our present work investigates whether starspots can produce an apparent lithium dispersion purely through surface temperature inhomogeneities, even in the absence of intrinsic abundance differences.}

In this study, we re-evaluate whether the observed lithium spread in NGC~2264 can be explained by the temperature effects of starspots. Using high-resolution ESPaDOnS spectra and 
ASAS-SN (All-Sky Automated Survey for Supernovae) photometry of weak-lined T~Tauri stars (WTTSs), we demonstrate that the EW(Li) varies with stellar rotation. We then construct a simple empirical model to quantify how starspot coverage influences the apparent EW(Li). 
Through this approach, we assess that much of the observed lithium dispersion in young clusters can be attributed to surface inhomogeneities rather than intrinsic depletion. 


The article is structured as follows. \S2 describes the data and the sample we selected. \S3 analyzes the connection between EW(Li) and V-band magnitude and establishes a relationship between the variability of lithium and starspots. \S4 presents a two-component photosphere model to reproduce the observed lithium spread. \S5 reviews and discusses the possible interpretation of the lithium spread. The conclusions of this study are summarized in \S6.

{Throughout this work, we treat the star as a two temperature photosphere, with $T_{\rm phot}$ as the warmer surface and $T_{\rm spot}$ as the cooler spot.  The photospheric temperature $T_{\rm phot}$ is assumed to be equivalent to the temperature measured from high-resolution optical spectra.  The real photosphere is likely much more complex, including a range of temperatures and faculae that would be seen as hot spots.}


\section{Observational Data}
\begin{figure}[!t]
\centering
\includegraphics[width=0.45\textwidth]{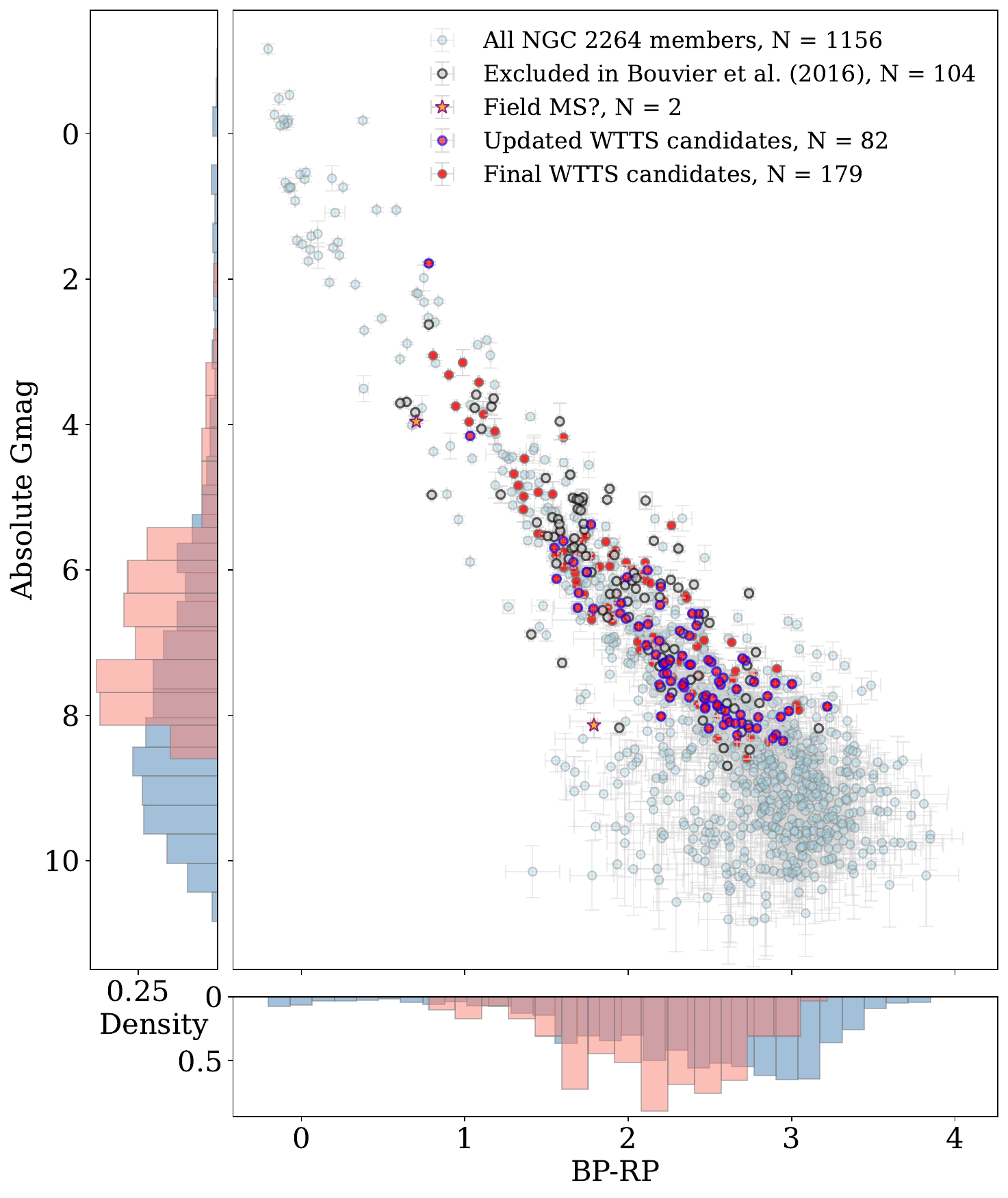}
\caption{Color-Magnitude diagram (CMD) of NGC~2264. The absolute G-band magnitude is calculated using the average cluster parallax. The grey points are all the members selected here from Gaia DR3, based on their proper motion and parallax (see Appendix A). The red points are WTTSs from \citet{bouvier2016gaia} that are consistent with membership.   The blue points are stars in \citet{bouvier2016gaia} excluded in work (low-confidence members or non-WTTSs).  The two orange stars are considered to be field stars based on their placement in the CMD, but cannot be excluded by Gaia DR3 astrometry. Extinction correction is not applied in this plot. The blue and red histograms show the distributions of all NGC~2264 members and the final WTTS candidates, respectively. }
\label{fig:NGC2264_CMD}
\end{figure}

\subsection{High resolution spectroscopic monitoring}
The demonstration of variability of lithium equivalent widths is based on a sample of 15 WTTSs selected from Zeeman Doppler Imaging (ZDI) campaigns, many from the Magnetic Topologies of Young Stars and the Survival of close-in giant Exoplanets program \citep[MaTYSSE, e.g.][]{Donati2014MaTYSSE}, using the ESPaDOnS instrument on the Canada-France-Hawaii Telescope (CFHT) \citep{donati2006espadons}. These campaigns obtained high signal-to-noise ratio (S/N), $R=65,000$ optical spectra across multiple stellar rotation periods, enabling a consistent set of Li measurements. The observations in our sample were conducted between 2009 and 2021. {Our collection is intended to demonstrate how spots modulate EW(Li) in young stars, rather than to provide a complete or unbiased sample. Stars in this sample come from different nearby regions, including Taurus, Lupus, TW Hydrae Association, and the Orion Nebula Cluster.}

The analysis of these spectra focuses on the Li\,\textsc{i} 6708\,\AA\ absorption line and the TiO molecular band near 7140\,\AA, both of which are sensitive to temperature inhomogeneities caused by cool starspots \citep[e.g.][]{oneal98}. To characterize the S/N of each spectrum, we coadd all spectra for each target to calculate a high S/N reference spectrum. Individual spectra are divided by this reference spectrum. We then calculate the S/N in the continuum regions (with weak absorption lines, e.g., 6701-6705\,\AA \,and 6711-6714\,\AA) as the ratio of the median flux to the standard deviation of the residuals. The median S/N for single exposures of all targets is $\sim 40$ in the lithium line region. All analysis in this paper is performed on individual exposures. {The typical measurement error of EW(Li) is $\sim$ 10\,m\text{\AA}.}

\subsection{Photometric monitoring with ASAS-SN}
Photometry of the WTTSs is obtained from the ASAS-SN light curves \citep{shappee2014man} and the ASAS-SN Variable Star Catalog \citep{jayasinghe2018asas}. {The typical measurement errors of V-band magnitude are $\sim0.02$}. In order to minimize the impact of long-term spot evolution, ASAS-SN photometry was selected to be as close in time as possible to the ESPaDOns spectra, although contemporaneous coverage is not available for all targets. The photometric data used in this work span the period from 2012 to 2018.

In Table~\ref{tab:wtts}, we list each of these sources with spectral types, photosphere temperatures from high-resolution spectra, rotation periods, median lithium equivalent widths, and median V-band magnitudes with their standard deviations, along with the date ranges for both the ESPaDOns spectra and the ASAS-SN photometry. The rotation periods are computed using the Lomb–Scargle algorithm \citep{Lomb1976,Scargle1982} through the NASA Exoplanet Archive Periodogram Service \citep{akeson2013nasa}. {Since several of our targets are strongly spotted, their listed spectral types measured from optical spectra are likely biased to hotter temperatures.}


\subsection{The membership and past Li measurements of NGC 2264}

\begin{table*}[!t]
\centering
\caption{Summary statistics for the NGC~2264 sample used in this work. We report the number of stars, the median [IQR] uncertainties in $T_{\rm phot}$ and EW(Li), and the observed dispersion in $\Delta{\rm EW(Li)}$ for the WTTS subsamples.}
\label{tab:ngc2264_summary}
\begin{tabular}{lcccc}
\toprule
Sample & $N$ & $\sigma_{T_{\rm phot}}$(K), Median[IQR]  & $\sigma_{\rm EW}$(m\AA), Median[IQR]  & $\mathrm{std}[\Delta{\rm EW(Li)}]$ (m\AA) \\
\midrule
Members & 407 & 57[53--67] & 5.1[3.0--8.4] &  \\
All WTTS & 179 & 57[53--62] & 5.2[3.0--8.6] &  39 \\
WTTS $\le3900$ K & 114 & 54[52--57] & 7.5[4.7--10.2] &  44  \\
$3900<$ K WTTS $\le 5000$ K & 55 & 67[62--72] & 3.7[2.5--4.7] &   31 \\
WTTS $\ge 5000$ K & 10 & 93[58--94] & 1.8[1.4--2.9] &  21  \\
\bottomrule
\end{tabular}
\end{table*}

{ 
We re-evaluate the spread in lithium equivalent widths of NGC 2264 published using 
measurements in the Gaia-ESO Survey Data Release\,5.1 \citep[GES DR\,5.1,][]{Gilmore2022ges,Randich2022ges,Worley2024ges}, obtained from the ESO Science Archive Facility\footnote{The GES DR\,5.1 catalog is derived from the ESO archive. DOI: \url{https://doi.org/10.18727/archive/25}}. The catalog provides stellar parameters and EW(Li) derived from the UVES spectra ($R=47000$, centered at $5800\,\text{\AA}$ with a spectral band of $2000\,\text{\AA}$) and the GIRAFFE spectra ($R=17000,\ 6470-6790\,\text{\AA}$) \citep{Lanzafame2015ges, Franciosini2022cog}. The photospheric temperatures in the Gaia-ESO survey were measured by template matching from multiple independent analysis nodes with a median error of 60\,K, while EW(Li)s were derived from the Arcetri node with a median error of 5\,m\AA.
}

\begin{figure}[!t]
\centering
\includegraphics[width=0.45\textwidth]{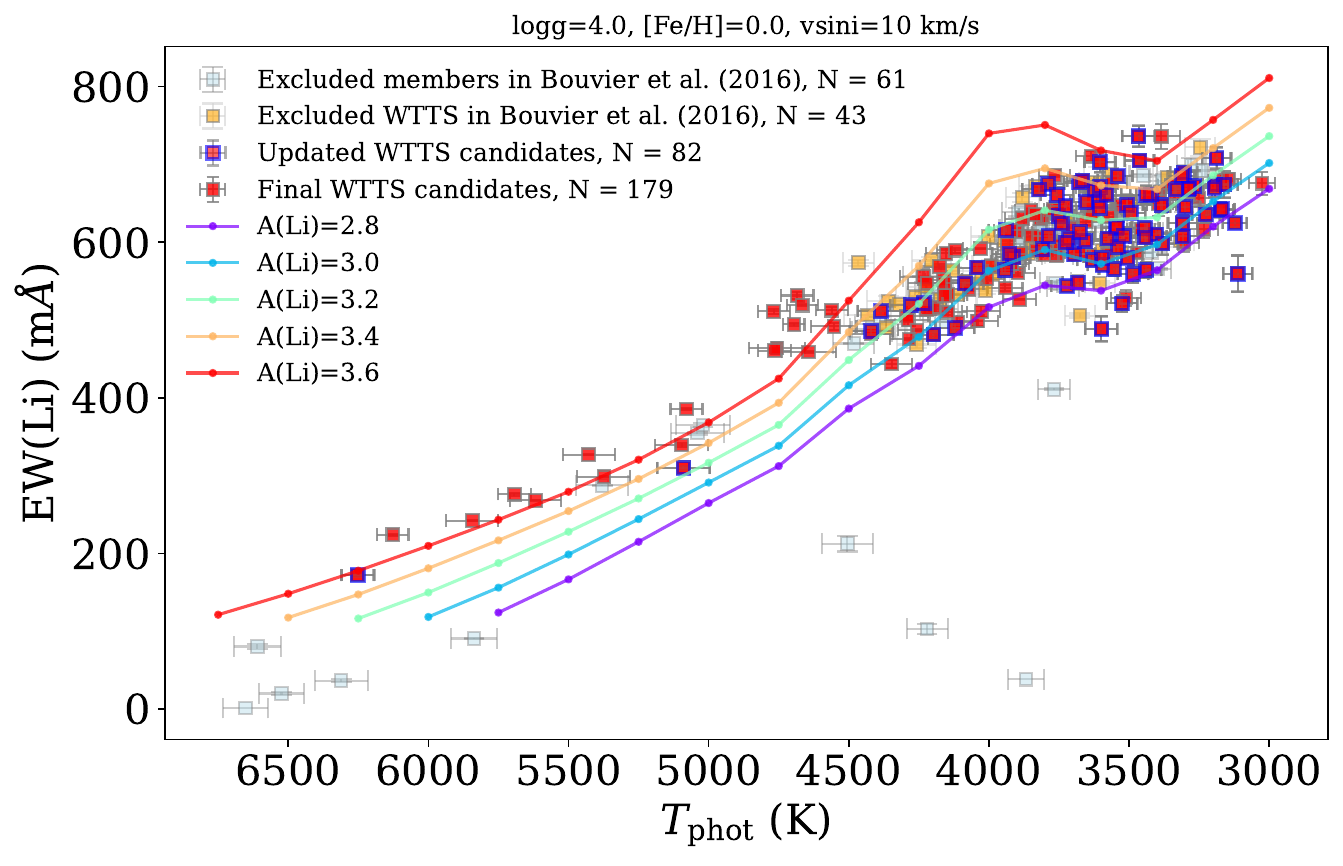}
\caption{Updated sample of NGC~2264 stars with Li measurements. Red points indicate WTTS members updated using GES DR5.1 and Gaia DR3. Red points with blue boxes mark updated members that are not included in \citet{bouvier2016gaia}. Orange points denote non-members excluded from \citet{bouvier2016gaia}, and grey points indicate excluded stars that are not WTTSs.}
\label{fig:li_spread_ngc2264}
\end{figure}

We reassess the cluster membership using high-precision proper motions and parallaxes from Gaia DR3 (Gaia Collaboration et al. \citeyear{gaia2016}, \citeyear{vallenari2023gaia}). 
Only very high confidence members are retained to ensure that the lithium spread originates within NGC~2264.
{The updated WTTS sample is based on the previous classification \citep{Venuti2018av} and is further refined using IR excess from J-W1 and J-W2 (using 2MASS photometry from \citealt{Skrutskie20062mass} and ALLWISE photometry from \citealt{Wright2010wise, Cutri2014allwise}) and H$\alpha$ accretion diagnostics from \citet{White2003ha}.
The detailed selection criteria are presented in Appendix~\ref{app:A}, and the updated NGC~2264 membership and Li dispersion are shown in Table~\ref{tab:ngc2264_summary}, Figure~\ref{fig:NGC2264_CMD} and Figure~\ref{fig:li_spread_ngc2264}. After applying these strict selection criteria, 1156 of the stars are retained as high-confidence members of NGC~2264, of which 179 high-confidence WTTSs that have EW(Li) and effective temperature measurements. Compared with data in \citet{bouvier2016gaia}, 
our sample includes 82 WTTSs that were not in the \citet{bouvier2016gaia} analysis and excludes 104 stars in their analysis.  Most of the 104 excluded stars are likely real members cut due to H$\alpha$ emission properties or strict astrometric criteria. Furthermore, the temperatures measured by GES DR5.1 are systematically higher than temperatures provided by \citet{bouvier2016gaia}.}

At $3900<T_{\rm phot}\le5000$\,K, the standard deviation of $\Delta{\rm EW(Li)}$ in the \citet{bouvier2016gaia} sample is dominated by four extremely Li-poor outliers, all of which are rejected by our selection. After excluding these four sources, the overall dispersions are nearly identical, 40 and 39\,m\AA, respectively, indicating that our membership and WTTS selection does not substantially change the characteristic Li spread.

\section{Variability of Lithium Equivalent Widths}
\label{sec:3}

{Our analysis of lithium variability is based on a sample of 15 WTTSs (described in \S~2.1) that serve as a demonstration of the role of spot modulation, even though they are not members of NGC 2264. These stars have multi-epoch high-resolution ESPaDOnS spectra and simultaneous photometry that allow us to characterize the relationship between EW(Li) and spot-modulated variability. We first use LkCa~4 as a detailed case study in \S~3.1, and then extend the analysis to the remaining 14 stars in \S~3.2. In \S~3.3, we quantify the variability amplitudes 
of EW(Li), TiO index, and V-band magnitude for the full 
sample and examine how these correlate with each another.}

{ 
We note that the \ion{Li}{1}\,6708\,\AA\ line is blended with the \ion{Fe}{1}\,6707.43\,\AA\ line. In our variability analysis, we measure the total equivalent width without deblending the \ion{Fe}{1} line contribution.}
{ 
The strength of the blend depends on temperature and line width \citep[e.g.][]{soderblom1993evolution,Franciosini2022cog}. Therefore, part of the variability of the unresolved 6708\,\AA\ feature may arise from blends rather than from \ion{Li}{1} alone. Below 3900\,K, the \ion{Fe}{1} line weakens and molecular blending becomes more significant \citep{randich2001ic}, which may further complicate the results in this work. Additionally, the Li line is dominated by the $^7$Li fine-structure doublet but also has possible contributions from the rarer $^6$Li isotope \citep{campbellwhite23}.}

\begin{figure*}[!t]
\centering
\includegraphics[width=0.9\textwidth]{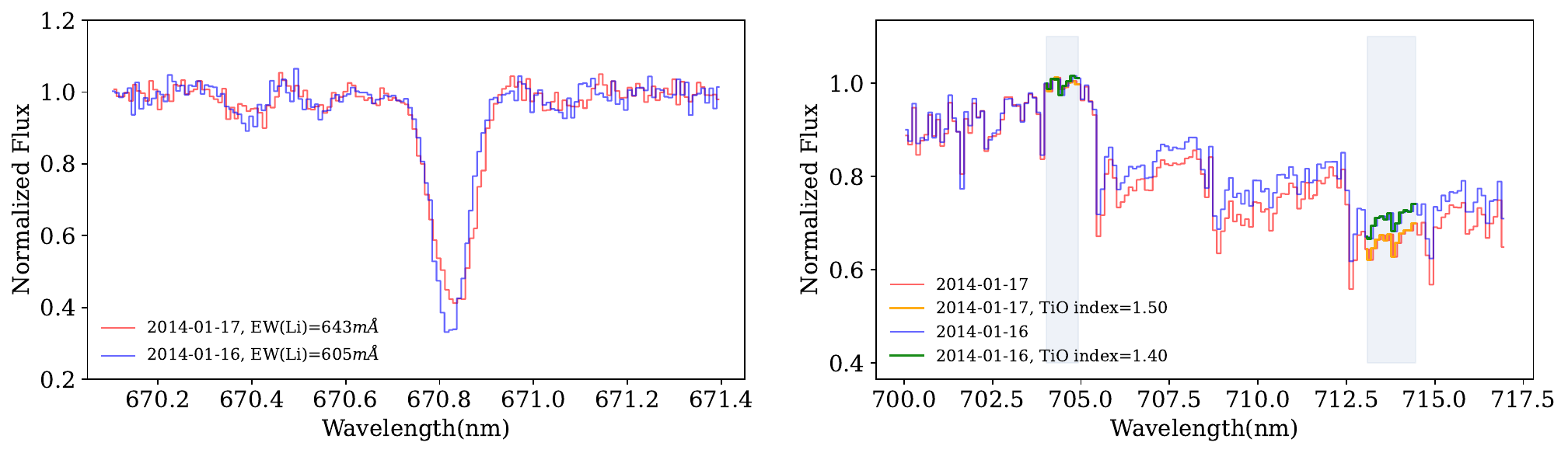}
\caption{Spectral features changes with the rotation of LkCa~4. Left panel: the Li 6708\AA\ absorption lines. The blue spectrum has larger EW(Li) than the orange one. The \ion{Li}{1} line appears as a single line but is the combination of four lines, doublets in $^6$Li and $^7$Li \citep{campbellwhite23}. Right panel: the TiO absorption band. The TiO index is defined as the ratio of the median flux in the continuum window (7040--7050\,\AA) to that in the band window (7130--7145\,\AA).}
\label{fig:Li_vs_TiO_extreme_lkca4}
\end{figure*}

\subsection{Demonstrating variability in spectra of LkCa~4}
\label{sec:3.1}

LkCa~4 is a well-studied WTTS with a large visible spot coverage that ranges from 67--83\% as the star rotates {\citep{gully2017placing,perez2023lkca4}}, making it an ideal target for testing the lithium-starspot connection. In this subsection, we use ESPaDOnS spectroscopy together with ASAS-SN photometry to show that the EW(Li) of LkCa~4 varies coherently with the stellar rotation. Our analysis demonstrates that  EW(Li) is correlated with both optical brightness and the TiO band depth, providing direct evidence that starspots modulate the observed lithium absorption.

We first measure the EW(Li) in each ESPaDOnS spectrum. The spectra are normalized in the continuum region around the lithium absorption line.  The equivalent width is then calculated by the integration of the Gaussian profile fit over a fixed wavelength interval (6701--6714\,\AA). The local continuum level is determined from adjacent line-free regions. The uncertainty of EW(Li) is estimated using the approximation from \citet{Cayrel_1988}:
\begin{align}
\delta EW \approx 1.5\,\frac{\sqrt{\Delta\lambda \cdot FWHM}}{S/N} 
\end{align}
where $FWHM$ is the full width at half maximum of the spectral line, $\Delta\lambda$ is the pixel size in angstroms, and $S/N$ is the signal-to-noise ratio.

\begin{figure*}[!t]
    \centering
    \subfigure{
    \begin{minipage}{0.48\textwidth}
        \includegraphics[width=\textwidth]{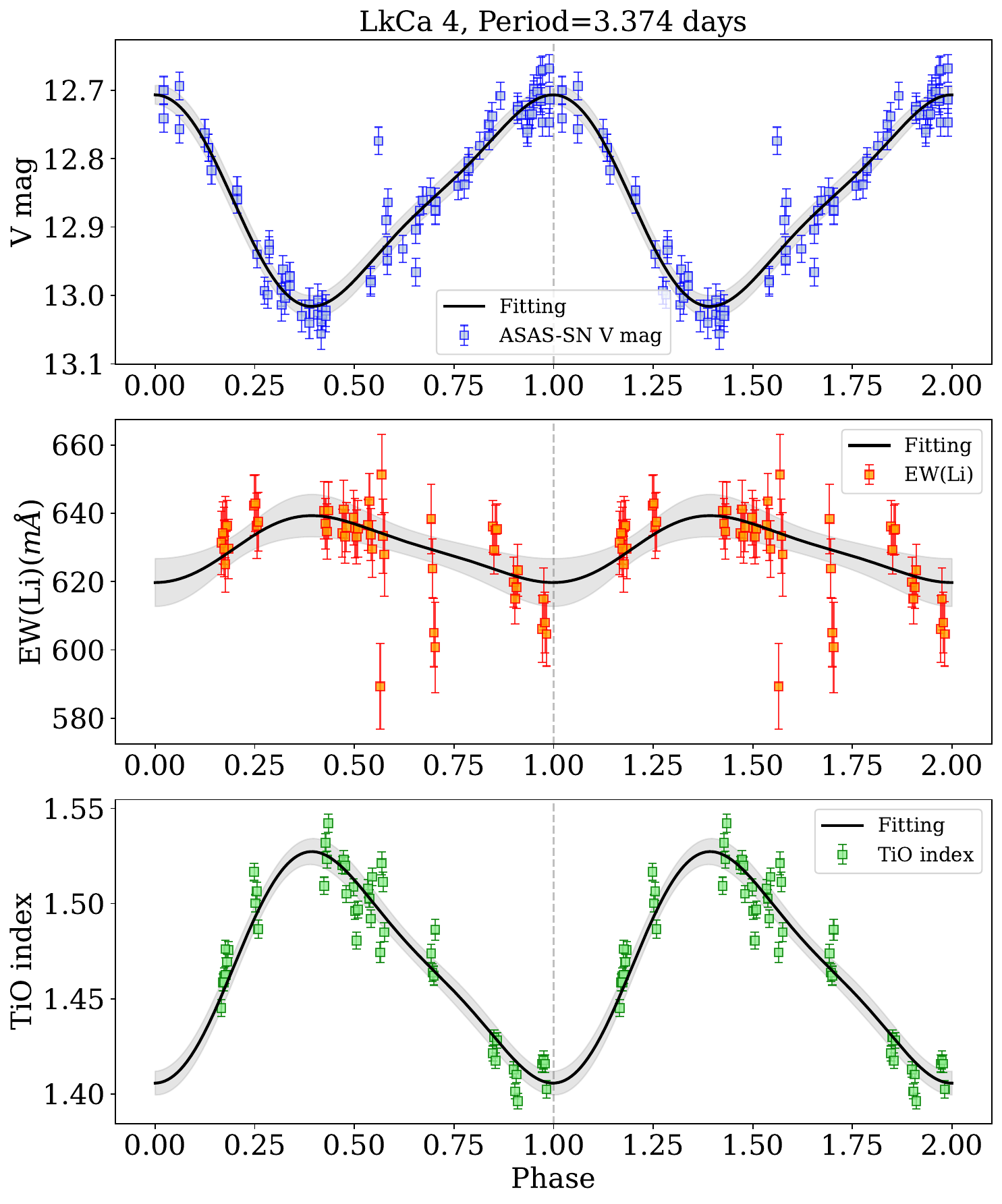}
        \label{fig:variability_compare_lkca4}
    \end{minipage}
    }
    \hfill
    \begin{minipage}{0.44\textwidth}
    
    \subfigure{
    \includegraphics[width=0.8\textwidth]{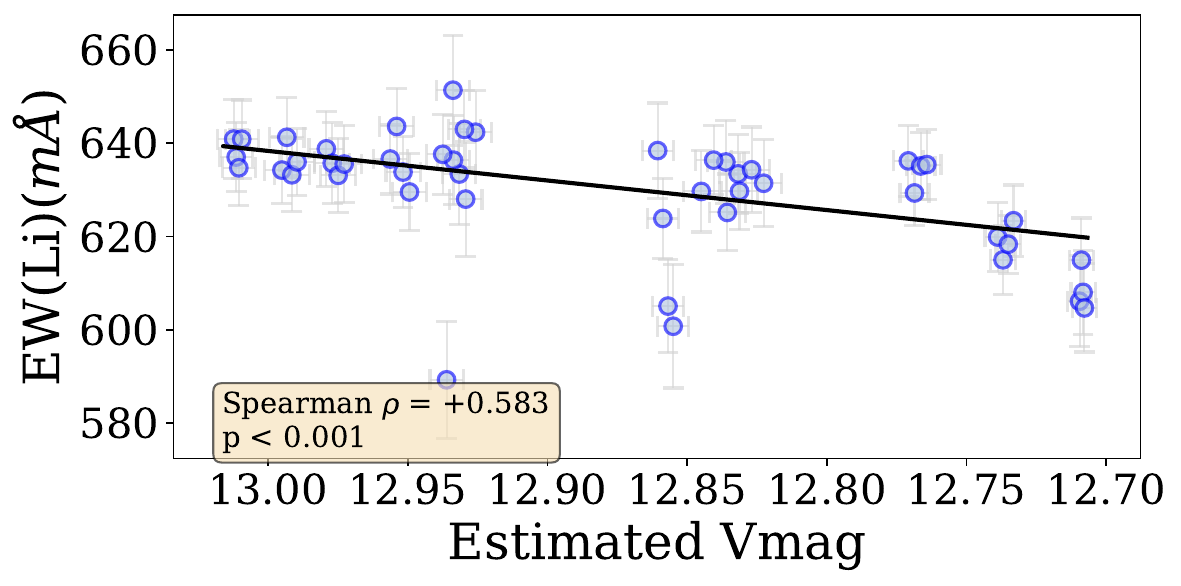}
        \label{fig:correlation_test_lkca4}
    }
    \subfigure{
        \includegraphics[width=0.8\textwidth]{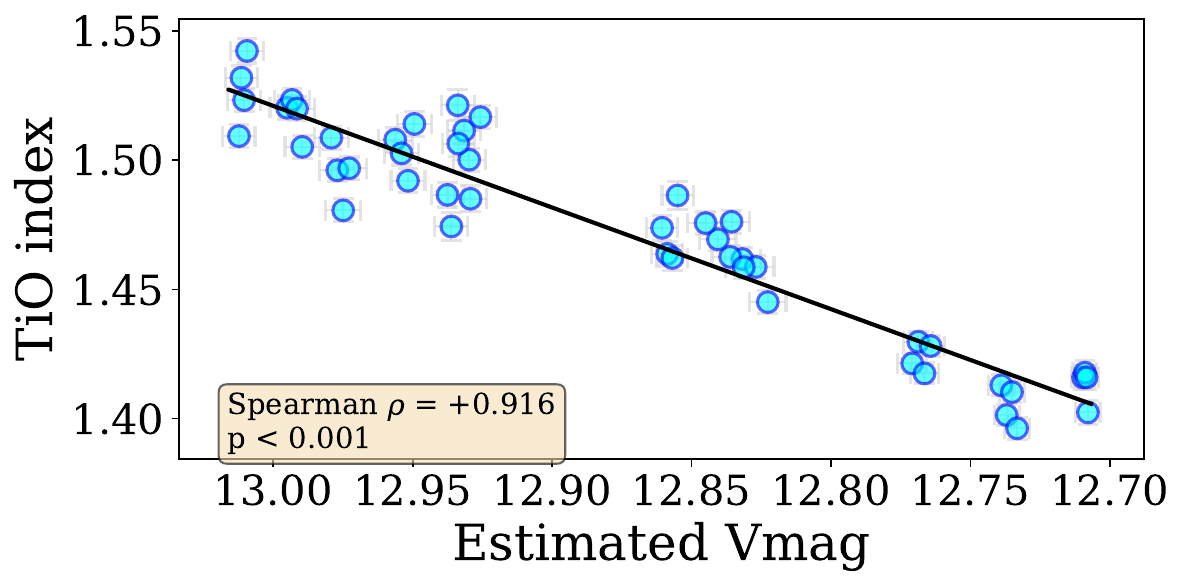}
        \label{fig:TiO_correlation_lkca4}
    }
    \subfigure{
        \includegraphics[width=0.8\textwidth]{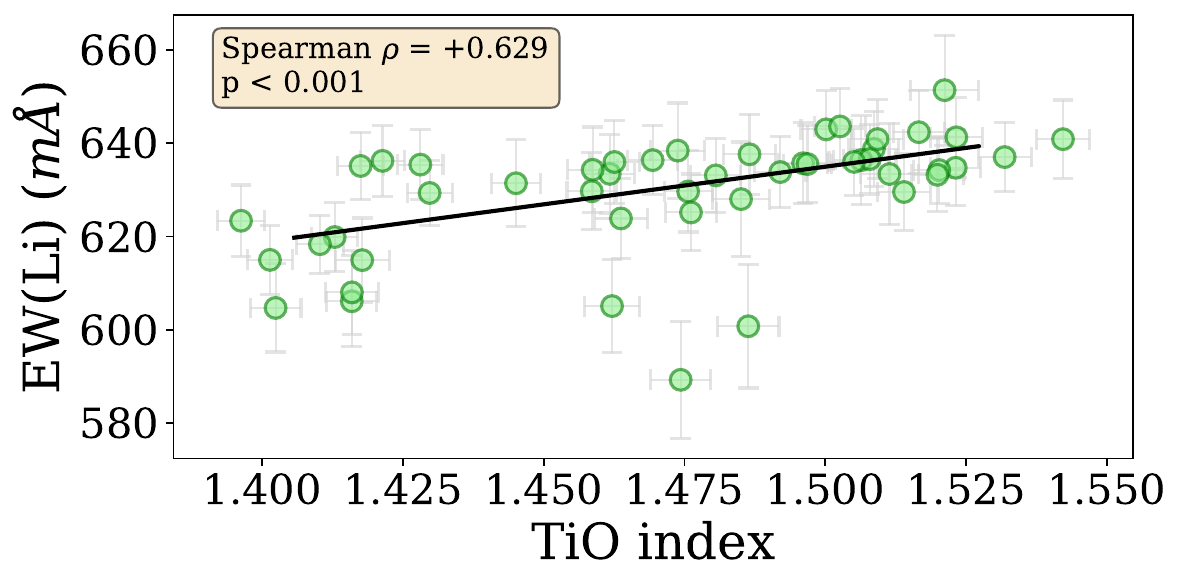}
        \label{fig:Li_vs_TiO_lkca4}
    }
    \end{minipage}
    
    \caption{(a) Left: Phased variations of starspot-related observables for LkCa~4. Blue, red and green symbols denote V-band magnitude, EW(Li) and TiO index; solid line is the best-fit Fourier series and shaded area the 3$\sigma$ fit uncertainty. (b) Top-right: EW(Li) vs V-band variation. (c) Middle-right: TiO index vs V-band variation. (d) Bottom-right: EW(Li) vs TiO index.}
    \label{fig:li_spot_lkca4}
\end{figure*}

To compare the temporal variations of EW(Li) with stellar surface activity, we select 75 ASAS-SN photometric points nearest in time to the spectroscopic observations (see Table~\ref{tab:wtts}). We then fit the phase-folded V-band magnitudes using a second-order Fourier series (see also \citealt{gully2017placing}), as
\begin{align}
f(\phi) = a_0 + &a_1 \cos(2\pi \phi) + b_1 \sin(2\pi \phi) \nonumber \\
+ &a_2 \cos(4\pi \phi) + b_2 \sin(4\pi \phi) 
\end{align}
where $\phi$ is the rotational phase, $a_0$ represents the mean level, and $a_1$, $b_1$, $a_2$, $b_2$ are the coefficients of the fundamental and first harmonic terms. The harmonic terms are added because starspot distributions that are inhomogeneous may produce light curves that are not perfect sinusoids \citep{grankin2008}. The uncertainties of the fitted light curves are estimated via Monte Carlo resampling of the data.
We then fix the fitted phase and only change the relative amplitude to fit the phase-folded EW(Li) and TiO index to 
compare different observables (V-band brightness, EW(Li), and TiO index) as functions of rotational phase. The fitting results are shown in Figure~\ref{fig:variability_compare_lkca4}.

Figure~\ref{fig:correlation_test_lkca4} shows that larger EW(Li) occurs when the star is fainter. The EW(Li) of LkCa~4 spans a range of $\sim600$-$650\,\text{m\AA}$ with typical uncertainties of $\sim10\,\text{m\AA}$, while the V-band magnitudes vary by $\sim 0.3$\,mag with uncertainties of $\sim0.005$\,mag. The V-band magnitude corresponding to each spectroscopic observation is estimated from the fitted light curve in Figure~\ref{fig:variability_compare_lkca4}. Spearman's rank correlation \citep{spearman1961proof} yields $\rho=+0.583$ with $p<0.001$ for EW(Li) versus V-band magnitude, indicating a strong correlation.

To further test the connection between EW(Li) and spots, we use the TiO~7140\,\AA\ band as an independent spectroscopic tracer of cool starspots
\citep[e.g.][]{neff1995absolute, gully2017placing}. 
We quantify the depth of the TiO-7140 band using a flux ratio index defined as
\begin{align}
\mathrm{TiO\,Index} = \frac{\langle I_\mathrm{cont} \rangle}{\langle I_\mathrm{band} \rangle} 
\end{align}
{where the median intensity of the continuum $\langle I_\mathrm{cont} \rangle$ and the TiO absorption band $\langle I_\mathrm{band} \rangle$ are measured within the wavelength ranges 7040--7050\,\AA\ and 7130--7145\,\AA, respectively. The TiO absorption feature is shown in Figure~\ref{fig:Li_vs_TiO_extreme_lkca4}.}

The EW(Li) is also correlated with the TiO index, as seen in Figure~\ref{fig:Li_vs_TiO_lkca4}, {which is consistent with the results in \citet{perez2023lkca4}.} For EW(Li) ranging from 620--640\,m\text{\AA}, the corresponding TiO index increases from 1.40 to 1.55, with a correlation ($\rho=+0.629$, $p<0.001$) supporting that spot coverage modulates the apparent EW(Li). Some dispersion, including outliers below $600$\,m\text{\AA}, may result from low spectrum quality, measurement uncertainties, {chromospheric flares,} or perhaps temporary changes of surface features.


\subsection{Measuring lithium variability in other young stars}
\label{sec:3.2}

{ 
In \S~\ref{sec:3.1} we established a correlation between EW(Li) and starspots for LkCa~4. 
In this subsection, we extend the analysis to 14 additional WTTSs (see Appendix~B for individual light curves and correlations). 
Most of these stars are analyzed using the same procedures {to measure Li strength from Gaussian profiles, TiO depth, and photometric variability from ASAs-SN.} 
{The \ion{Li}{1} equivalent widths for ROX~39, TAP~26, TWA~6, and V410~Tau are obtained by direct integration over a narrower interval (6705--6711\,\AA) because their lines are not well described by a Gaussian profile.}
For V410~Tau we use contemporaneous \(V\)-band photometry from the Crimean Astronomical Observatory \citep{yu2019v410}, because its light‑curve amplitude and morphology evolve on yearly timescales \citep{hamb2019v410}. 
For RX~J1608.0-3857 and RX~J1609.5-3850, only non-contemporaneous ASAS-SN data are available, which may introduce systematic offsets \citep[see seasonal variability of WTTSs spots in][]{grankin2008}.}

{ 
The EW(Li) correlates with the V-band magnitude and the TiO index in the same sense as for LkCa~4. Fainter V-band magnitudes and larger TiO indices correspond to larger EW(Li)s. The TiO index correlates more tightly with the V-band magnitude than EW(Li) does for 14 of the 15 stars in our sample, whereas the TiO--EW(Li) correlations become weaker and more scattered. This may result from the different SNRs of the Li line and the TiO band in the spectra. We use Spearman rank correlation between EW(Li) and V-band magnitude to quantify the relation. All \(\rho\) and \(p\) values are listed in Table~\ref{tab:wtts_correlation}. 
Five targets, LkCa~7, RX~J1609.5-3850, TWA~25, TWA~6, and V819~Tau, show a strong correlation (\(\rho > 0.6\) with \(p < 0.001\)).  
Three targets, ROXs~45F, RX~J1608.0-3857, and V830~Tau, have \(0.5 < \rho < 0.6\) with \(p < 0.001\), similar to the correlation probability for LkCa~4: . 
The remaining 6 stars show weak or no correlation, mostly because of low variability in V-band of $<0.1$\,mag and in EW(Li) below 15\,m\AA\. 
One exception, TAP~45 (K6, P=9.587\,d), shows a weak negative correlation, such that EW(Li) decreases as the star becomes fainter and TiO absorption strengthens, although this is likely due to small variation in V-band magnitude of $\sim$0.04\,mag and a median measurement error of EW(Li) (5.9\,m$\text{\AA}$) larger than its standard deviation (5.1\,m$\text{\AA}$).}

{ 
Cool stars show strong EW(Li) -- TiO index correlations, while hotter stars barely show any correlation.
The M-type stars in our sample, including LkCa~7, RX~J1609.5-3850, and TWA~25, show strong TiO absorption tightly related with EW(Li). In contrast, stars with spectral types earlier than K5 are too hot to show any TiO band absorption, and the index variations can be taken as typical measurement uncertainty for stars of similar spectral type and S/N (for the K3 V410~Tau, ${\rm TiO_{std}}\sim$0.006; for the K4 Par~1379, ${\rm TiO_{std}}\sim$0.005; for the K5 ROX~39, ${\rm TiO_{std}}\sim$0.007). 
}

\begin{figure*}[!t]
\centering
\includegraphics[width=0.9\textwidth]{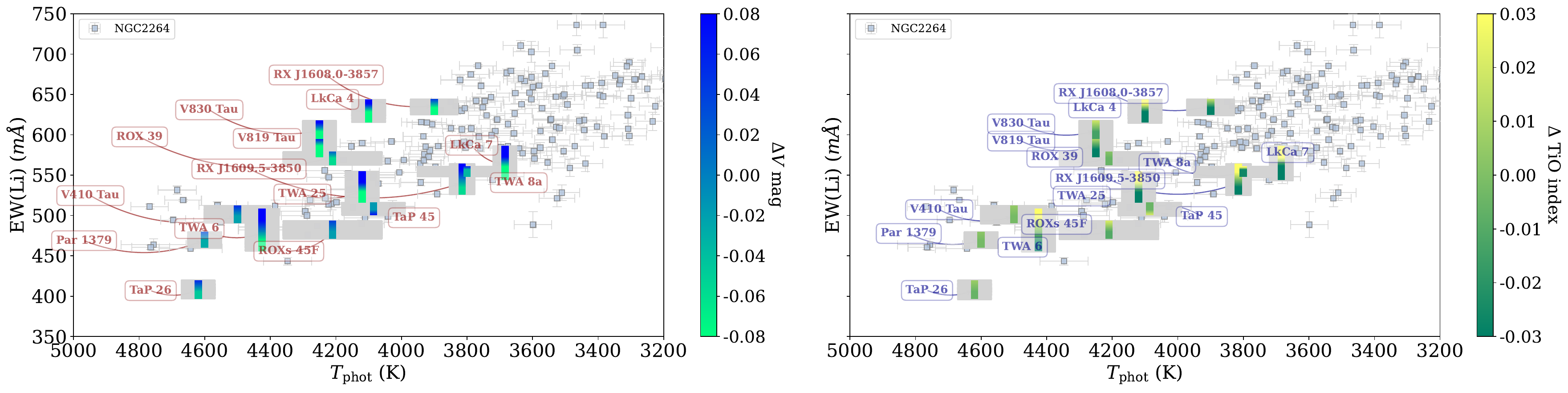}
\caption{Place the EW(Li) variation of WTTSs on the updated NGC~2264 EW(Li) vs.\ $T_{\rm phot}$ plot. The color bars indicate the changes of V-band magnitudes (left panel) and TiO indices (right panel) relative to the median. {Horizontal grey bars show the uncertainties in temperatures. Photospheric temperatures and their uncertainties are adopted from MaTYSSE papers, where they were derived from high-resolution optical ESPaDOnS spectra. For stars lacking such spectroscopic determinations, $T_{\rm phot}$s are estimated from spectral types with typical uncertainties in \citet{herczeg2014optical}: 1 subclass for G0--K8, 0.5 subclass for K8--M0.5, and 0.2 subclass for later than M0, and propagated these uncertainties to the temperatures.}}
\label{fig:EWLi_vs_Teff_vmag_tio_moreTT}
\end{figure*}

This sample of 15 WTTSs provides a demonstration of the role of spots in EW(Li) measurements, despite a lack of variability and spot fraction information for the NGC 2264 sample.
Compared with the NGC~2264 WTTS sample, the 15 WTTS
stars span a larger range of ages in different regions, i.e., from the few-Myr populations in Taurus, Lupus, and Orion to the $\sim$10\,Myr TW Hydrae Association. They are also systematically weighted toward warmer stars: the median [IQR] adopted $T_{\rm phot}$ is 4210 [3990--4340]\,K, compared with 3740 [3506--4098]\,K for NGC~2264, and only 4/15 versus 114/179 stars have $T_{\rm phot}\le3900$\,K. Nevertheless, the distribution of rotation period (median, [IQR] = 3.37 [2.15--5.30]\,d) is basically consistent with that of the NGC~2264 sample (median, [IQR] 4.25 [2.17--6.55]\,d). So the behavior of the 15 WTTS stars is reasonable to establish a relation between lithium and starspots.
In Figure~\ref{fig:EWLi_vs_Teff_vmag_tio_moreTT}, the observed EW(Li) variations are overlaid on the updated EW(Li) versus $T_{\rm phot}$ diagram in NGC~2264. The variations of EW(Li)s are basically located within the lithium scatter of the NGC~2264 population.  
For individual stars as they rotate, their brighter magnitudes correspond to a lower EW(Li) and smaller TiO index. Therefore, the lithium spread in NGC~2264 can be explained, at least in part, by the difference in the spot coverage of the member stars.

\subsection{Lithium Variability and Photometric Variability}
\label{sec:3.3}

\begin{figure*}[!t]
    \centering
    \includegraphics[width=0.85\textwidth]{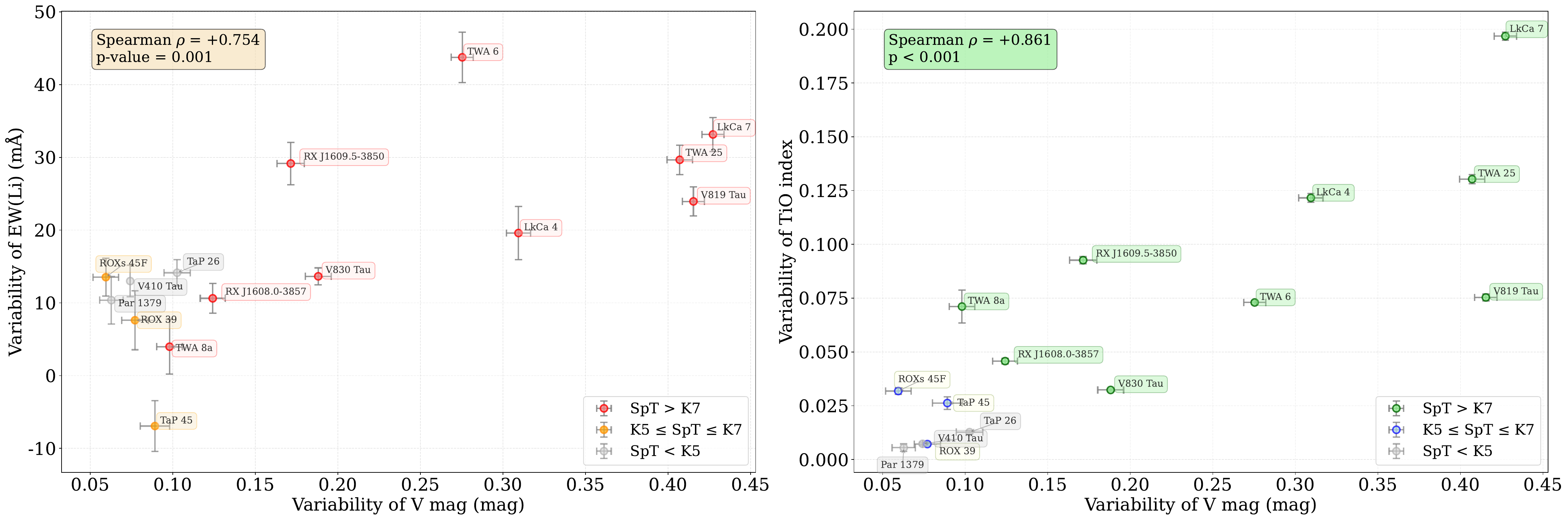}
    \caption{Left: The variability of EW(Li) versus V-band magnitude. Right: The variability of TiO index versus V-band magnitude. For each WTTS, their variabilities are quantified as the peak-to-peak amplitude of the fitting curve. The uncertainties are derived from 1000 Monte Carlo realizations. Stars in bright colors (red or green) have spectral type later than $K7$; stars in moderate colors (yellow or blue) have $K5\le SpT \le K7$; stars in grey colors have spectral type earlier than $K5$, which barely retain TiO absorption band.}
    \label{fig:variability_ewli_vmag_tio}
\end{figure*}

{In this section, we demonstrate that the variability of EW(Li), TiO index, and V-band magnitude are correlated with each other.}
For each star, we define the variability as the peak-to-peak amplitude (maximum minus minimum) of the Fourier-fit periodic curves  (see \S\ref{sec:3.1}). The uncertainties of the variability are estimated through 1000 Monte Carlo realizations. For each realization, we use parameters resampled from the multivariate normal distribution defined by the best-fit values and their covariance matrix, and recompute the peak-to-peak amplitude. Then the dispersion of the amplitudes is adopted as the uncertainty, reflecting uncertainties in the fit rather than in individual measurements. The results are shown in Figure~\ref{fig:variability_ewli_vmag_tio}.

The variability of EW(Li) and the TiO depth generally increases with the variability of the V-band magnitude (see Figure~\ref{fig:variability_ewli_vmag_tio}). The Spearman's rank correlation indicate that the correlation of the variability of the TiO index versus V-band magnitude ($\rho=+0.861$, $p<0.001$) is stronger than the variability of the EW(Li) versus V-band magnitude ($\rho=+0.750$, $p=0.001$). 
{In both panels, warmer stars (SpT $\leq$ K5) show small variability. The variations of their EW(Li) are typically $\lesssim20$\,m\AA, V-band amplitudes $\lesssim0.1$\,mag, and TiO index $\lesssim0.05$. Within this temperature range, the variability amplitudes do not exhibit clear systematic trends because molecular bands form only at temperatures below the photospheric temperatures of warm stars.}

A few sources stand out from this general pattern. TWA~6 (M0) has small V-band variability ($\lesssim 0.3$~mag) yet its EW(Li) varies by $\sim40$~m\AA. This may reflect strong spot coverage, including high-latitude and asymmetric starspots, or compensation from plages, as seen in its ZDI study \cite{Hill2019twa}. TAP~45 exhibits an anti-correlation between EW(Li) and spots, as discussed in \S~\ref{sec:3.2}. This behavior is likely dominated by measurement scatter. Hotter stars (SpT $<$ K5) nearly show zero variability in the TiO index because their atmospheres lack molecules, but still have certain EW(Li) variability, even larger than some cooler stars.

\section{Reproducing lithium spreads with starspots}
\label{sec:4}

{ 
We have demonstrated that EW(Li) varies coherently with spot coverage on rotational timescales, suggesting that surface temperature inhomogeneities 
contribute to the observed lithium spread in NGC~2264. We first quantify this contribution using an empirical two-temperature model and examine the distribution of spot coverage required to reproduce the observations. We then use the curves of growth from \citealp{Franciosini2022cog} to compare the predicted contributions of starspots, surface gravity, and age.} 

\subsection{An Empirical 2-Temperature Model}

{We construct a simple empirical model to interpret the observed EW(Li) spread as a surface-temperature effect caused by starspots. The model is based on our updated EW(Li)--$T_{\rm phot}$ relation in NGC~2264.
The dispersion of the EW(Li) residuals,
\begin{align}
    \Delta{\rm EW(Li)} \equiv {\rm EW(Li)}-{\rm EW(Li)}_{\rm fit}(T_{\rm phot}) 
\end{align}
where ${\rm EW(Li)}_{\rm fit} $ is the piecewise linear fit to the EW(Li)-$T_{\rm phot}$ relation, increases systematically toward lower temperatures. The standard deviation of $\Delta{\rm EW(Li)}$ is 21\,m\AA\ for $T_{\rm phot}\ge5000$\,K, 31\,m\AA\ for $3900<T_{\rm phot}\le5000$\,K, and 44\,m\AA\ for $T_{\rm phot}\le3900$\,K (Table~\ref{tab:ngc2264_summary}).
We exclude the few stars hotter than 5000\,K from most of the following analysis. 
We then define the unspotted region as the lower envelope of the EW(Li)--$T_{\rm phot}$ distribution. A linear least-squares fit to the median of the lowest 20\% of EW(Li) values in each temperature bin provides an empirical relation for the baseline of our model, assumed for our purposes to be unspotted.}


{ 
We model each star as a combination of a warm photosphere and a cool spot component \citep[e.g.][]{somers2015rotation, gully2017placing, perez2025spectral}. The spot temperature is assumed to satisfy $\log(T_{\rm spot})=\log(T_{\rm phot})-0.12$, similar to previous studies \citep[e.g.][]{jackson2014effect,somers2020spots}. This prescription is simplified but sufficient for our purpose. For a given $T_{\rm phot}$, we compute the apparent EW(Li) by combining the equivalent widths of the two components, weighted by their fluxes and the spot filling factor $f_{\rm spot}$ :
\begin{align}
EW_{\rm out}=\frac{EW_{\rm phot}F_{\rm phot}(1-f_{\rm spot})+EW_{\rm spot}F_{\rm spot}f_{\rm spot}}{F_{\rm phot}(1-f_{\rm spot})+F_{\rm spot}f_{\rm spot}} 
\end{align}
Here $F_{\rm phot}$ and $F_{\rm spot}$ are taken from BT-Settl model spectra \citep{allard2014exploring} with $\log g=4.0$ and $[{\rm M/H}]=0.0$ (The median metallicity of NGC~2264 members is −0.060 $\pm$ 0.006, \citealp{spina2017metal}).
This approach is analogous to the analytical model of \citet{2005XiongSpot}, but uses an empirical lower-envelope fit instead of theoretical curves of growth to estimate EW(Li). Varying $f_{\rm spot}$ then produces a grid of model predictions in the $T_{\rm phot}$--$f_{\rm spot}$ plane.
}

\subsection{EW(Li) Predictions from the Two-Temperature Model}
\label{sec:4.2}
\begin{figure}[!t]
\centering
\includegraphics[width=0.45\textwidth]{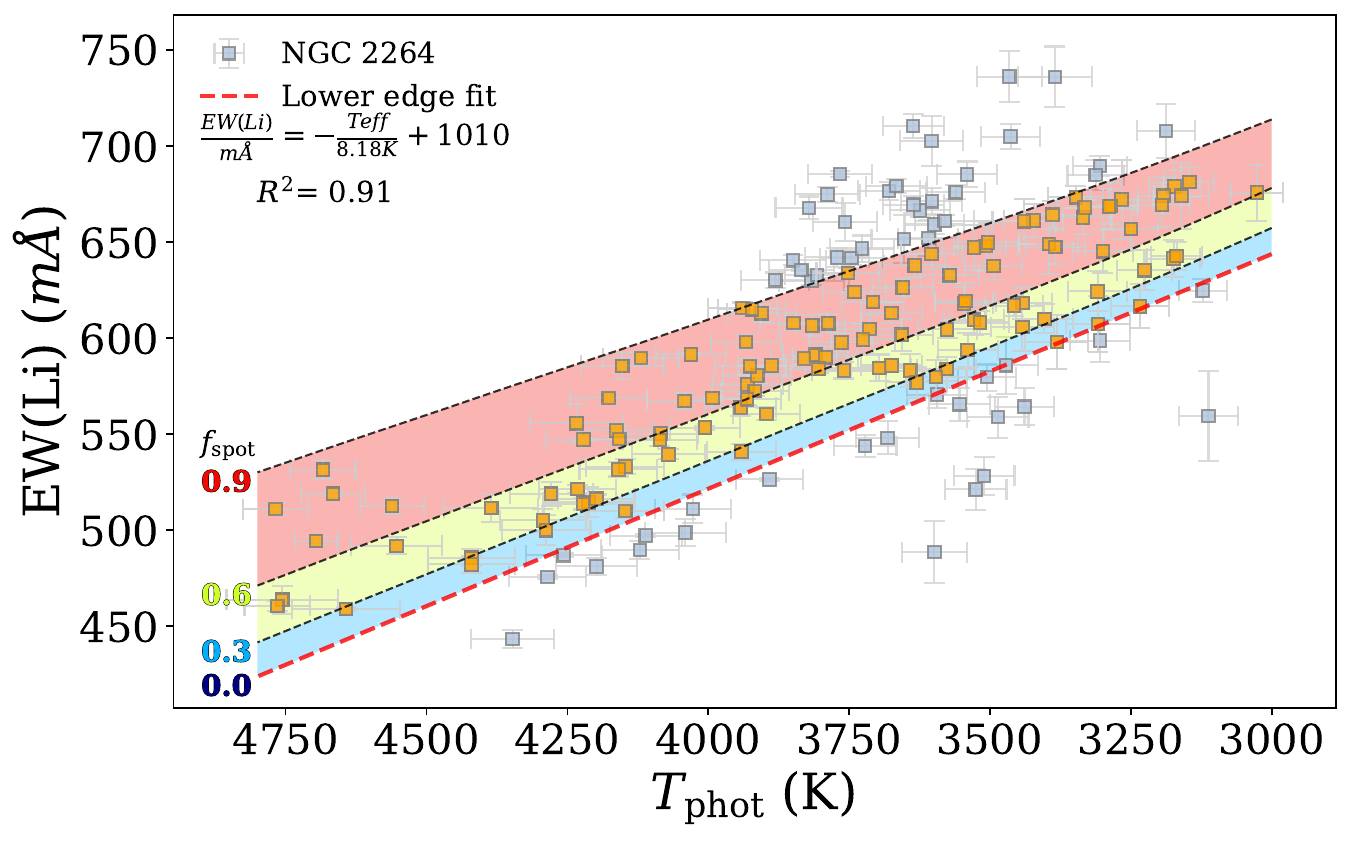}
\caption{Empirical two-temperature spot model. The grey points show the updated WTTSs in NGC~2264. The red dashed line is the linear fit to the lower envelope of the EW(Li)--$T_{\rm phot}$ relation for stars with $3000<T_{\rm phot}<5000$~K, which we adopt as the unspotted baseline. The colored shaded bands show the predicted EW(Li) for increasing spot filling factors, assuming a cool spot component with $\log(T_{\rm spot})=\log(T_{\rm phot})-0.12$. 68\% stars lie within the model predicted region. Larger temperature contrasts produce a larger EW(Li) spread (see Appendix~\ref{APP:C}).}
\label{fig:spot_model}
\end{figure}

The model provides the apparent EW(Li) as a function of photospheric temperature $T_{\rm phot}$ and spot coverage $f_{\rm spot}$. In Figure~\ref{fig:spot_model}, the output EW(Li) from the model yields a grid of model predictions across $f_{\rm spot}=0.0$ to $0.9$. For unspotted stars ($f_{\rm spot} = 0.0$), the linear fit relationship follows 
\begin{align}
    \rm EW(Li)/m\text{\AA} = -T_{\rm phot}/8.18\,K + 1010 
\end{align}
with a high correlation coefficient ($R^2 \sim 0.91$). 

When starspots are introduced, an increasing spot coverage leads to a larger EW(Li) at fixed $T_{\rm phot}$, producing a vertical spread similar to that observed in NGC~2264. 
{The Li spread increases from $\sim$70\,m\AA\ at $T_{\rm phot}=3000$\,K to $\sim$110\,m\AA\ at $T_{\rm phot}=5000$\,K for $f_{\rm spot}=0$--0.9, which is because we assume a fixed ratio between the unspotted photospheric and spot temperatures. The vertical displacement of the model covers 68\% of the WTTSs in NGC~2264. The outliers are predominantly stars cooler than 3900\,K, which exhibit a larger lithium dispersion (Table~\ref{tab:ngc2264_summary}). The model reproduces 85\% of the stars with $3900<T_{\rm phot}\le5000$\,K, but only 60\% of those with $T_{\rm phot}\le3900$\,K.}
At higher $T_{\rm phot}$, the absolute temperature contrast between the photosphere and spots is larger, producing a greater enhancement of EW(Li).

The model also predicts that spot coverages below $\sim$0.6 produce only a modest EW(Li) spread, while coverages in the 0.6--0.9 range yield a substantially larger vertical dispersion.  Such high spot coverages are consistent with spectroscopic measurements from \citet{perez2025spectral} for $\sim$4000 K, but their spot coverage measurements for cooler stars are smaller and would not produce a spread.
This non-linear increase implies that reproducing the full observed envelope of lithium strengths in NGC~2264 requires a large population with high spot coverages. Typical spot coverages of $\gtrsim50\%$ have been inferred for young, active pre-main sequence stars \citep[e.g.][]{somers2015rotation,Fang2016spot}, and even higher fractions may be common \citep[e.g.][]{perez2025spectral}.


We also test the influence of different photosphere-spot temperature contrasts. Figure~\ref{fig:li_spot_model_Tspot_dex} shows the spot models with different photosphere-spot temperature contrasts, i.e., $\log T_{\rm phot} - \log T_{\rm spot} = 0.06, \ 0.12, \ {\rm and} \ 0.18 $, and $T_{\rm phot} - T_{\rm spot} =600,\ 900, \ {\rm and} \ 1200$\,K. A larger photosphere-spot temperature contrast further enhances the lithium dispersion, as also discussed in \citet{somers2015rotation} in terms of radius inflation and lithium depletion.

{ 
The change in EW(Li) in our model is consistent with the changes seen in EW(Li) due to rotational modulation of the visible spot coverage (Section~\ref{sec:3}). For example, for LkCa~4, spectroscopic and photometric analysis from \citet{gully2017placing} indicates that the photosphere has a warm component of 4100\,K and a spot component of 2750\,K, with a spot covering fraction that varies from 67--83\% with rotation.  For these parameters, the EW(Li) should increase by 34\,m$\text{\AA}$. The spot parameters ($T_{\rm phot}=4100-4400\,K,\ T_{\rm spot}=3060\,K,\ f_{\rm spot}=0.77-0.94$) derived from \citet{perez2023lkca4} yield a EW(Li) enhancement of $\sim$41--57\,m$\text{\AA}$. To compare, the observed EW(Li) of LkCa~4 ranges from 590--650\,m$\text{\AA}$. However, the absolute EW(Li) values for LkCa~4 and several other stars in our WTTS sample are higher than those predicted by our model.
}

\subsection{Spot distribution required for the lithium spread}
\label{sec:4.3}

{For $f_{\rm spot}=0$--$0.9$, the model spans a similar EW(Li) range to the observations, but 54 of 169 stars (32\%) lie outside the predicted envelope, with 31 above and 23 below it (see Appendix~\ref{APP:C} for other spot-temperature prescriptions). Stars with EW(Li) excesses (larger than the predicted $f_{\rm spot}=0.9$ line) are all cooler than 3900~K.}

To assess how much of the observed EW(Li) spread can be reproduced, we generate a synthetic sample by randomly drawing spot coverages from a Gaussian distribution centered at $50\%$ with a dispersion of $20\%$. In this case, the standard deviation of the simulated EW(Li) residuals is $19\,\text{m\AA}$, substantially smaller than the observed dispersion of $42\,\text{m\AA}$.

\begin{figure*}[t]
\centering
\includegraphics[width=0.9\textwidth]{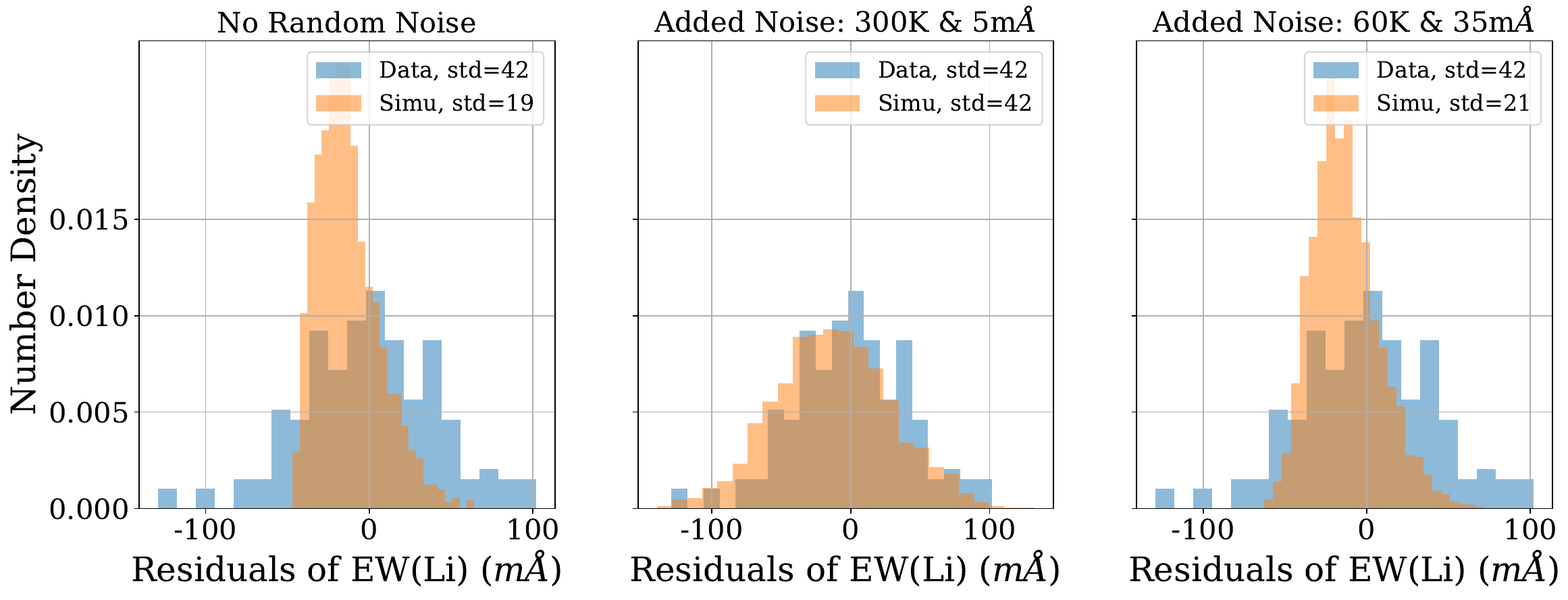}
\caption{Comparison of the EW(Li) residual ($\Delta$EW(Li)) distribution between NGC~2264 (3000-5000\,K, in blue) and synthetic data (in orange), where $\Delta$EW(Li) is defined as the difference between the observed EW(Li) and the value predicted by linear fitted EW(Li) at the same $T_{\rm phot}$. 
The left panel shows synthetic data with spot coverages  drawn from a Gaussian distribution centered at 50\% with a dispersion of 20\%. The middle and right panels further include Gaussian uncertainties, $\sigma(T_{\rm phot}, {\rm EWLi)}\sim\mathcal{N}(300\,{\rm K},\,5{\rm m\text{\AA}})$ and
$\sigma(T_{\rm phot}, {\rm EWLi)}\sim\mathcal{N}(60\,{\rm K},\,35{\rm m\text{\AA}})$, respectively.}
\label{fig:EWLi_Spot_model_fspot_gaussian_hist_comparison_50_20_0-0.9}
\end{figure*}

{ 
Since part of the observed lithium spread may arise from measurement uncertainties, we incorporate observational errors into the model. We add random Gaussian noise of $60\,\mathrm{K}$ in $T_{\rm phot}$ and $5\,\text{m\AA}$ in EW(Li), following the median measuring errors in our updated NGC~2264 sample. The standard deviation of the EW(Li) residuals only increases to $21\,\text{m\AA}$ when the temperature and Li uncertainties are applied. To fully reproduce the observed lithium spread, we need to add Gaussian uncertainties of either $300$\,K in temperature or $35$\,m\AA\ in EW(Li) (see Figure~\ref{fig:EWLi_Spot_model_fspot_gaussian_hist_comparison_50_20_0-0.9}), much larger than the median errors listed in the GES archive.  Our starspot model combined with observational uncertainties cannot fully account for the lithium dispersion in NGC~2264.
}

\subsection{Possible Contributions of Age and Gravity to the Spread in Li Equivalent Widths}

{ 
Our analysis above shows that the starspot models can explain much of the lithium spread in NGC~2264. However, the residual scatter is not fully captured by the model, especially for stars lower than 3900\,K. The remaining scatter may arise from 
an intrinsic age spread among cluster members or differences in surface gravity. 
In this subsection, we examine whether age and surface gravity could also contribute to the observed dispersion. We find that both surface gravity and stellar age show no correlation with EW(Li) residuals, though models predict that they can produce measurable EW(Li) dispersions (summarized in
Table~\ref{tab:contributions}). 
}

\begin{figure*}[t] 
    \centering
    \subfigure{
    \begin{minipage}{0.5\textwidth}
        \includegraphics[width=\textwidth]{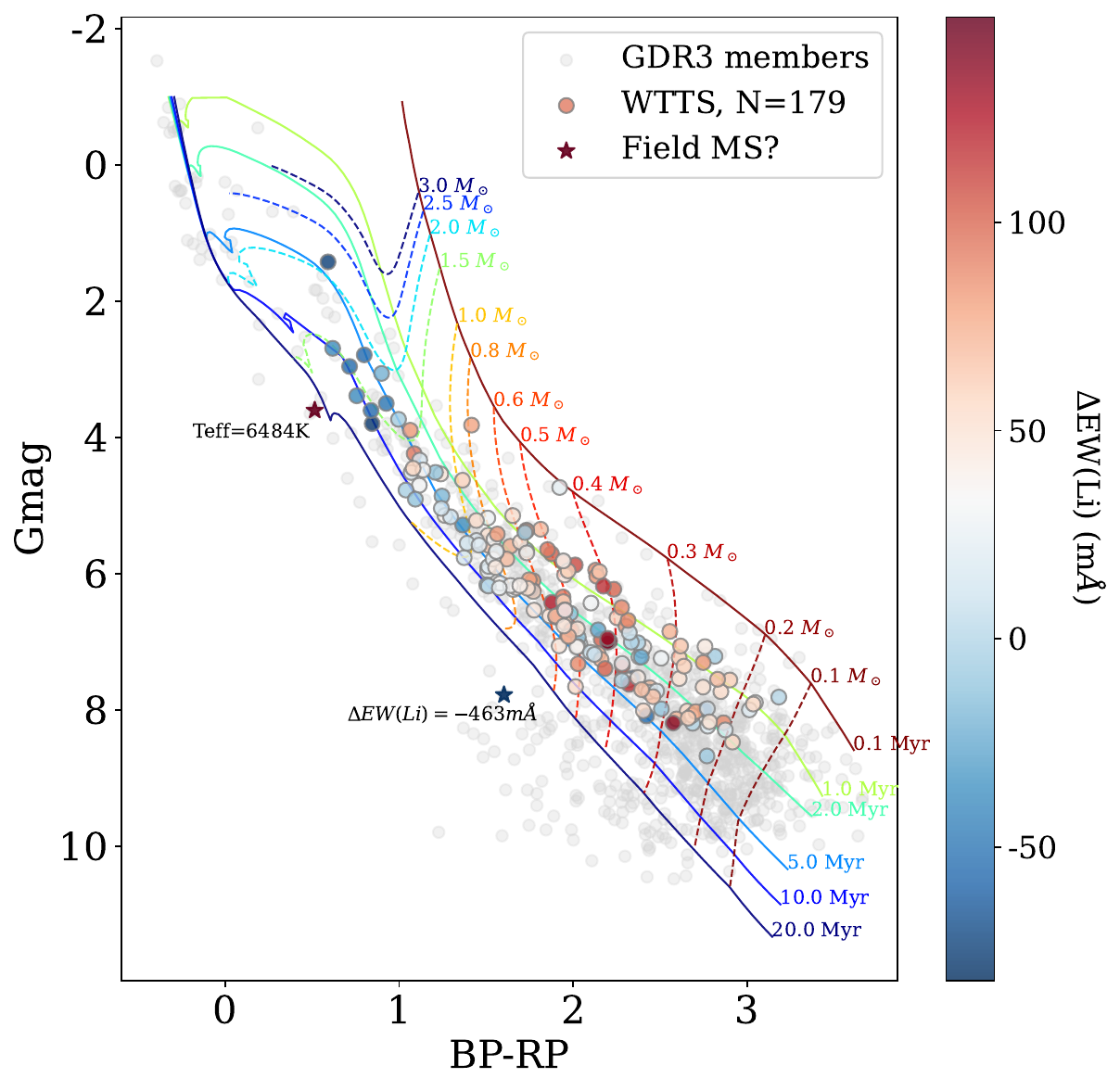}
        \label{fig:NGC2264_CMD_iso}
    \end{minipage}
    }
    \hfill
    \begin{minipage}{0.45\textwidth}
    \subfigure{
        \includegraphics[width=\textwidth]{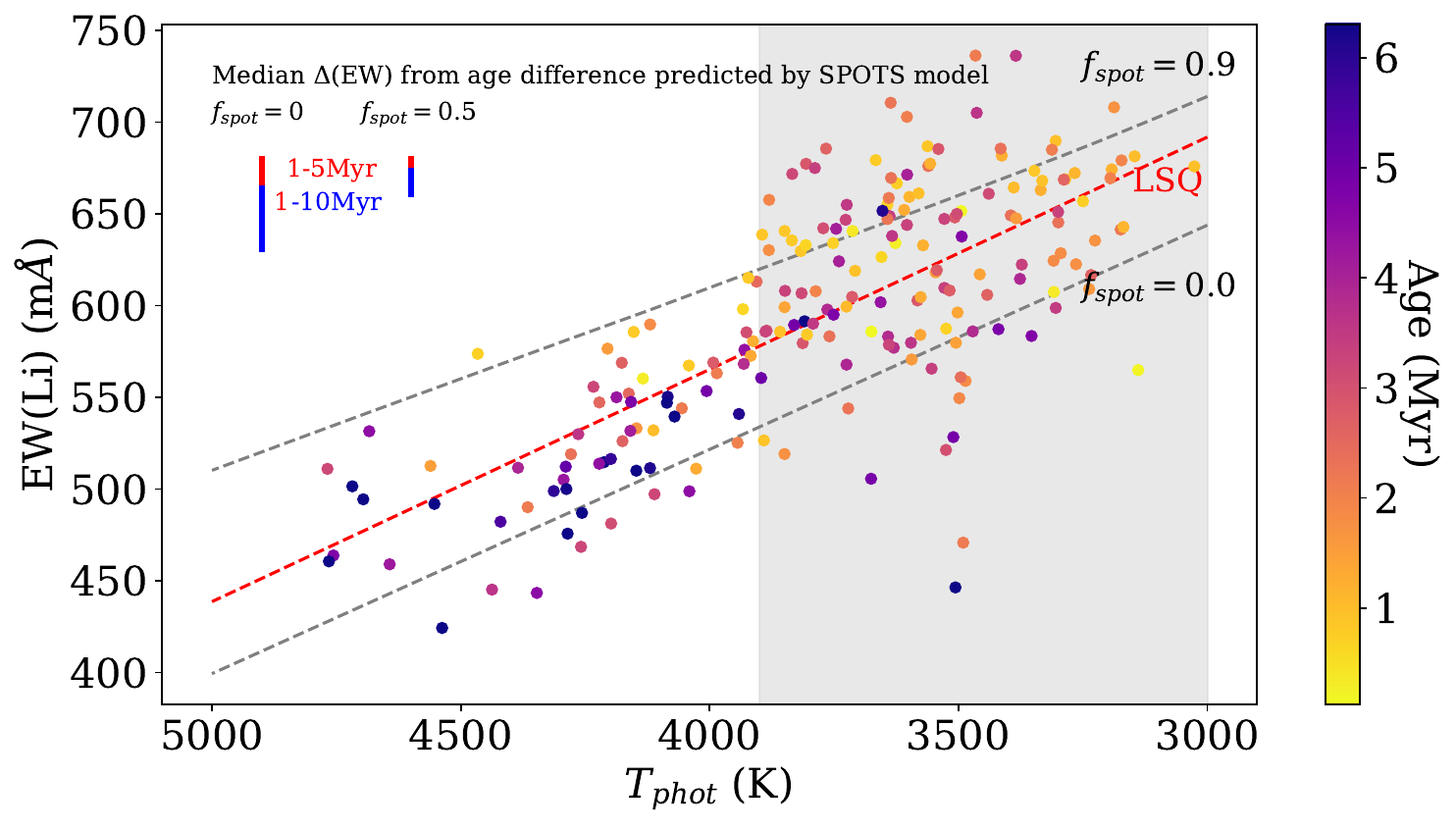}
        \label{fig:EWLi_Teff_G-K_age}
    }
    \subfigure{
        \includegraphics[width=\textwidth]{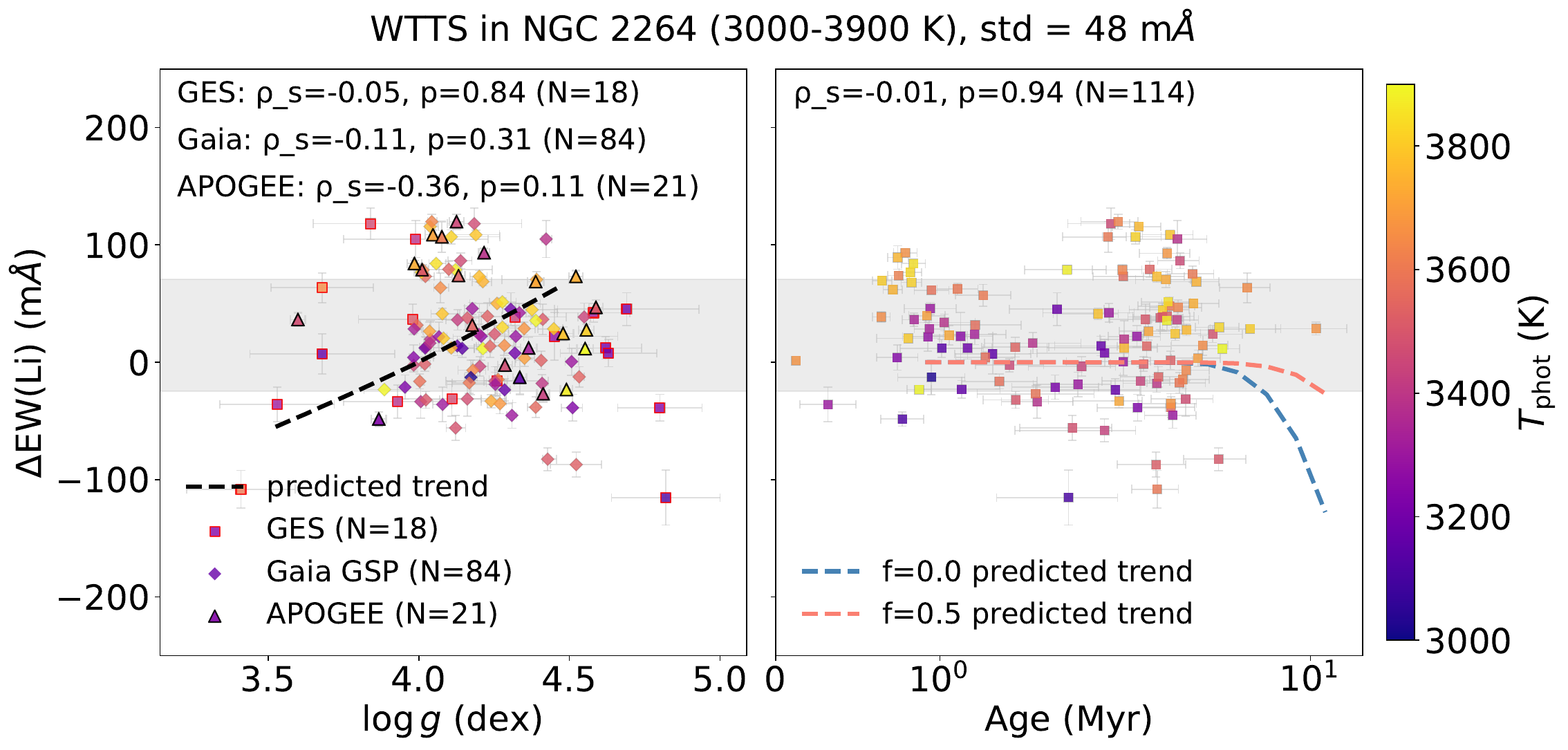}
        \label{fig:EWLi_age_cut3500_4200K_av_corrected}
    }
    \end{minipage}
    \caption{Left: Extinction-corrected CMD of NGC~2264 members using $A_V$ from \citet{Venuti2018av}, with MIST isochrones \citep{dotter2016mesa, choi2016mesa} overlaid as colored dashed lines. Top right: EW(Li) versus $T_{\rm phot}$, with grey dashed curves showing starspot model predictions for $f_{\rm spot}=0$ and $0.9$, and the red dashed line a linear fit over 3000--5000\,K. Point colors indicate MIST ages. Vertical segments show the median EW(Li) spread predicted by the SPOTS models \citep{somers2020spots} for age spreads of 1--5\,Myr (red) and 1--10\,Myr (blue). The grey shaded region marks stars with $3000<T_{\rm phot}<3900$\,K with larger Li spread. Bottom right: EW(Li) residuals versus $\log g$ (left panel) and age (right panel) for the $3000<T_{\rm phot}<3900$\,K subsample.    
    }
    \label{fig:li_spread_age}
\end{figure*}

\vspace{3mm}

{ 
\subsubsection{Expectations from spectral models}
}

{ 
        We calculate synthetic EW(Li) distributions using the curves of growth in \citet{Franciosini2022cog} to compare the effects of starspots to the expected effects from stellar age and surface gravity. We adopt [Fe/H]$=0$ throughout, consistent with the solar metallicity assumed in the preceding starspot model. We set $v\sin i=10$\,km\,s$^{-1}$, which is close to the median value of the GES measurement 13\,km\,s$^{-1}$ for the NGC~2264 WTTS sample and is available in the grid of the curves of growth. We repeat the calculations for A(Li)$=3.0$ and 3.6. 

We estimate the contributions of spots, age, and surface gravity using a Monte Carlo approach. For each effect we draw 10,000 samples from a Gaussian distribution while holding all other parameters fixed. At each $T_{\rm phot}$ we subtract the median synthetic EW(Li) and compute the standard deviation of the residuals, then binning them into the three temperature ranges set in Table~\ref{tab:ngc2264_summary}.

 Results from the curves of growth predict that spots, age, and surface gravity each produce certain EW(Li) dispersion (see Table~\ref{tab:contributions}). 
 For the effect of starspots, the spot coverage of the sample follows $f_{\rm spot}\sim\mathcal{N}(0.5,0.2)$, truncated to $[0,0.9]$. The curves of growth yield dispersions of 28--34\,m\AA\ below 5000\,K, which are systematically larger than the 16--20\,m\AA\ obtained from the empirical model in Section~\ref{sec:4.3}.

}

\vspace{3mm}
{ 
    \noindent\textit{Predictions for age:}
    The sample follows $\log(\mathrm{age/yr})\sim\mathcal{N}(6.40,0.34)$, truncated to $[5.1,7.1]$. We interpolate the Li depletion factor $(\mathrm{Li/Li}_0)$ from the SPOTS models \citep{somers2020spots} and convert the Li abundance into EW(Li). In the unspotted model, the Li depletion starts at around 5\,Myr at $T_{\rm phot}\sim 3700$--4500\,K, yielding EW(Li) dispersions of 21--37\,m\AA\ among 3000--5000\,K. 
    The $f_{\rm spot}=0.5$ spotted model delays the onset of depletion to 8--10\,Myr, and reduces the dispersion to 3--8\,m\AA. However, as seen in Figure \ref{fig:li_spread_age}, the Li dispersion from the estimated ages should be negligible.  The model dispersions likely overestimate the number of older stars in the distribution.
    
}

\vspace{3mm}
{ 
    \noindent\textit{Predictions for gravity:}      
    For the effect of surface gravity, the sample follows $\log g\sim\mathcal{N}(4.0,0.3)$ and is truncated to $[3.5,4.5]$, consistent with measured gravities and expectations for the age.  
    The predicted dispersion has its largest impact at low temperatures and depends on the assumed A(Li): for A(Li)$=3.6$, gravity produces 32\,m\AA\ below 3900\,K and 20\,m\AA\ between 3900 and 5000\,K, while for A(Li)$=3.0$ the corresponding values decrease to 22 and 8\,m\AA\ (Table~\ref{tab:contributions}). Gravity differences could in principle contribute substantially to the lithium spread of the coolest stars. Above 5000\,K, gravity produces only 1--2\,m\AA, much smaller than the observed 21\,m\AA.}

{ 
\subsubsection{Nondetection of correlations with age and gravity}
   }


{ 
    \noindent\textit{Noncorrelation with age:} To investigate whether an age spread among cluster members can contribute to the Li dispersion, we estimate individual stellar ages by comparing the extinction-corrected BP-RP versus G color-magnitude diagram with the MESA Isochrones and Stellar Tracks \citep[MIST,][]{dotter2016mesa,choi2016mesa}, as shown in Figure~\ref{fig:NGC2264_CMD_iso}. We adopt the extinction measurements of \citet{Venuti2018av}, who estimated $A_V$ by comparing the observed colors with the intrinsic colors of \citet{Covey2007sed}. For stars without individual extinction measurements, we adopt the median $A_V$ of NGC~2264. The resulting ages have a mean log(Age/yr) of 6.40 (2.5\,Myr) with a standard deviation of 0.34 (range from 5.09 to 7.08, i.e., 0.1 to 12\,Myr). These estimates do not include corrections for starspots, which if uncorrected biases color-magnitude diagram estimates to older ages and adds scatter \citep[e.g.][]{somers2015rotation}.
    }

{ 
    We find no significant empirical relation between the stellar ages and the EW(Li) residuals. In the upper right panel of Figure~\ref{fig:li_spread_age}, stars in 3000--3900\,K seem to have younger estimated ages, when compared to stars with higher temperatures. Within the 3000-3900\,K temperature range and after correcting for the correlation between temperature and EW(Li), no trend appears between EW(Li) residual and the stellar age ($\rho=-0.01$ with $p=0.94$, shown in the bottom right of Figure~\ref{fig:li_spread_age}). 
    The sample seems to separate into younger and older populations at approximately 1.5 Myr, the $\Delta$EW(Li) distributions of the two groups show no significant trend, with the Kolmogorov-Smirnov (KS) test result of $D= 0.17$ and $p=0.44$. The effect of the age spread on the Li dispersion cannot be excluded but is unlikely to be the primary source of the observed dispersion in a heavily spotted population. We also caution that the age spread derived from the CMD is not necessarily a real age spread within the cluster members. Although we have minimized the contamination of field stars from Gaia astrometry, other effects such as unresolved binaries, incorrect extinctions, and starspots can still mimic an age spread \citep[e.g.][]{Preibisch1999usco,Hartigan_Kenyon_2001,Venuti2018av, somers2015rotation}. Confirming intrinsic age differences still requires independent age diagnostics.
}

{ 
    \noindent\textit{Noncorrelation with gravity:}
    We do not detect any correlation between $\log g$ and EW(Li) residual in the available gravity measurements. Within $3000<T_{\rm phot}<3900$\,K, 18 stars have GES surface gravities measurement, but no correlation is found between $\log g$ and $\Delta{\rm EW(Li)}$ ($\rho=-0.05$ with $p=0.84$). The Gaia DR3 GSP-Phot gravities, inferred for 84 stars by forward modelling of the BP/RP spectra, parallax, and $G$ magnitude against PARSEC isochrones \citep{Andrae2023gsp}, also show no correlation with $\rho=-0.11$ and $p=0.31$. The APOGEE DR17 ASPCAP gravities, derived for 21 stars by spectral fitting of the H-band spectra against MARCS model atmospheres \citep{garcia2016apo,Majewski2017apo, Abdurro2022apo}, also show no significant correlation with $\rho=-0.36$ and $p=0.11$. 
}


\subsection{Summarizing contributions to the Lithium spread}

{ 
   
    
    Table~\ref{tab:contributions} summarizes the observed scatter in EW(Li) and the contributions from modeling different contributions.  
        The observed correlations between EW(Li), V-band magnitude, and the TiO band (Section~\ref{sec:3}) provide direct observational evidence that changes in visible spot coverage can modify EW(Li).  When combined with spots, the expected dispersion in surface gravity, age, and empirical uncertainties could entirely reproduce the dispersion in EW(Li).  However, no corresponding empirical relation is detected   either surface gravity or inferred age.

    If caused by real differences in Li abundance, the dispersion in EW(Li) has a median scatter of 0.09 dex and maximum differences of 0.37 dex.  While large Li abundance differences are not expected between objects within the same cluster, some minor fluctuations could add to the spread. The curve of growth of A(Li)$=3.6$ reproduces the observed EW(Li)--$T_{\rm phot}$ relation above 5000\,K, where the hotter stars are expected to be nearly unspotted. However, cooler stars have EW(Li) that are better reproduced by A(Li)$=2.8-3.2$ (see Figure~\ref{fig:li_spread_ngc2264}), the range discussed for NGC~2264 by \citet{bouvier2016gaia}.  This temperature dependence is likely caused by systematic differences and not real differences in abundance.
}



\begin{table*}[!t]
     \centering
    \caption{{Li dispersion (standard deviation of $\Delta{\rm EW(Li)}$ in m\AA) predicted by the curves of growth.}}
    \label{tab:contributions}

    \small
    \setlength{\tabcolsep}{12pt}

    \begin{tabular}{lcccccccccc}
    \toprule
    & & \multicolumn{9}{c}{Model contributions from} \\
    \cmidrule{3-11}
    $T_{\rm phot}$(K) & Observed$^{a}$ & Errors$^{b}$
      & \multicolumn{2}{c}{Spots$^{c}$}
      & \multicolumn{2}{c}{$\log g^{d}$}
      & \multicolumn{2}{c}{Age, 0 spot$^{e}$}
      & \multicolumn{2}{c}{Age, 51\% spots$^{f}$} \\
    \cmidrule(lr){4-5} \cmidrule(lr){6-7} \cmidrule(lr){8-9} \cmidrule(lr){10-11}
    $A({\rm Li})$ & -- & -- & 3.0 & 3.6 & 3.0 & 3.6 & 3.0 & 3.6 & 3.0 & 3.6 \\
    \midrule
    $\le3900$ & 44 & 8 & 30 & 28 & 22 & 32 & 30 & 37 & 3 & 8 \\
    $3900$--$5000$ & 31 & 9 & 33 & 34 & 8 & 20 & 21 & 31 & 3 & 4 \\
    $>5000$ & 21 & 13 & -- & -- & 1 & 2 & 2 & 1 & 0 & 0 \\
    \bottomrule
    \end{tabular}

    \small
    Notes.\ (a) Observed EW(Li) dispersion per $T_{\rm phot}$ bin.
    (b) Dispersion caused by $T_{\rm phot}$ and EW(Li) measurement uncertainties (Table~\ref{tab:ngc2264_summary}), via Monte Carlo perturbation.
    (c) Dispersion from spot coverage, $f_{\rm spot}\sim\mathcal{N}(0.5,0.2)$ truncated to $[0.0,0.9]$, using flux-weighted curves of growth with $\log T_{\rm phot}-\log T_{\rm spot}=0.12$.
    (d) Dispersion from $\log g\sim\mathcal{N}(4.0,0.3)$ truncated to $[3.5,4.5]$, at fixed $T_{\rm phot}$, [Fe/H]$=0$, $v\sin i=10$\,km\,s$^{-1}$.
    (e)--(f) Dispersion from age spread, $\log(\mathrm{age/yr})\sim\mathcal{N}(6.40,0.34)$ truncated to $[5.1,7.1]$, using SPOTS depletion models \citep{somers2020spots} with $f_{\rm spot}=0$ (e) or $0.5$ (f), at fixed $\log g=4.0$, [Fe/H]$=0$, $v\sin i=10$\,km\,s$^{-1}$.
\end{table*}

\section{Discussion}

{ 
In this study, we find that for the 15 WTTSs in our sample, EW(Li), photometric brightness, and TiO absorption vary coherently with stellar rotation, especially in cooler stars. Our empirical starspots model suggests that at least part of the lithium dispersion in NGC~2264 arises from spot-induced surface temperature inhomogeneities rather than intrinsic abundance differences alone. In the following subsections, we compare our findings with previous work and discuss their implications for lithium scatter and age determinations.
}

\subsection{Comparisons to previous work on Li variability}



Previous studies have reached different conclusions regarding the role of starspots in driving the lithium dispersion, likely reflecting differences in stellar samples and activity regimes. Early work established that starspots can modulate the observed lithium line strength in some stars.  \citet{robinson1986variation} first demonstrated that EW(Li) varies with rotational phase in two active stars, AB~Dor and PZ~Tel. 
\citet{king2000lithium} found in the Pleiades a correlation between the difference in measured Pleiades Li abundance and stellar activity, as measured from the Ca\,\textsc{ii} infrared triplet flux. These analyses establish that stellar activity may play an important part in causing the lithium spread in Pleiades and that the Li-rotation relation may be an appearance of the Li-activity relation. \citet{barrado2016seven} found a correlation between rotation period and stellar activity traced by photometric variability and X-ray emission in Pleiades, again supporting that magnetic activity should play a role in the EW(Li).

Other studies that reported weak or absent lithium modulation caused by starspots generally targeted hotter stars and are not in contradiction with our results. For instance, although EW(Li) variability in V410~Tau has been reported \citep{fernandez1998spectroscopy}, \citet{basri1991v410} found no clear relation between EW(Li) and spot coverage, but it was based on only four data points. In our work, we find weak EW(Li) variability and only a marginal correlation with rotation. The weak modulation likely reflects the relatively high photospheric temperature of V410~Tau (K3), at which the spot-to-photosphere contrast is smaller and the effect on EW(Li) is reduced. Cooler stars in our sample show stronger variability and correlations. Likewise, \citet{pallavicini1993effects} found no significant rotational modulation of the lithium line in several Pleiades stars and concluded that starspots are unlikely to be the dominant driver of the lithium dispersion. Their targets are older ($\sim125$\,Myr) and relatively hot (>4700\,K), with V-band variability below $\sim0.2$\,mag and EW(Li) below $\sim300$\,m\AA. Stars in similar temperature (>5000\,K) and EW(Li) (<400\,m\AA) regime also show less lithium scatter in NGC~2264. Our findings therefore extend rather than conflict with this earlier study.  Together these studies demonstrate that lithium variability associated with starspots is more pronounced in cooler stars (see \S~\ref{sec:3.3}).

\subsection{Implications for Lithium Scatter and Age Determinations}


Correctly interpreting the lithium spread requires understanding how long starspot-related effects dominate the observations, the temperatures of the spotted photosphere, and at what evolutionary stage intrinsic lithium abundance differences must be considered. As stars evolve, magnetic activity and spot coverage are expected to decline statistically \citep[e.g.][]{Morris2020spotage,Nichols2020spotage}, although the trend depends on mass and rotation and is not necessarily monotonic for all low-mass stars. If the lithium dispersion is always dominated by starspots, it would therefore be expected to decrease with cluster age. However, the lithium spread is found to increase with cluster age from around 10\,Myr, peak at around 50\,Myr. and then decrease \citep{jackson2025growth}. While the temperature effect caused by starspots can be a major explanation for the lithium spread in the youngest clusters, intrinsic differences in lithium abundance should be taken into account once lithium burning begins.


Young stars likely have higher spot coverage than is historically assumed. Our empirical modeling of NGC~2264 assumes a mean spot coverage of $\sim0.5$ with Gaussian distribution in cluster members. These values are consistent with recent spectroscopic modeling, yielding median spot fractions of $\sim0.5$ for 1--35\,Myr WTTSs \citep{perez2025spectral} and $\sim0.25$ for the Pleiades \citep{cao2022spot,Fang2016spot}. These values are higher than obtained from ZDI modeling, but ZDI may sometimes systematically underestimate spot coverage. For instance, in our test case of LkCa~4, 
the ZDI analysis yielded a spot covering fraction of $\sim0.25$ \citep{Donati2014MaTYSSE}, while spectral modeling and the large photometric variability suggest $0.67-0.83$ in \citealt{gully2017placing} and $0.77-0.94$ in \citealt{perez2023lkca4}). To reproduce the lithium dispersion, \citet{jackson2025growth} find that spot coverages would need to be about twice those measured in clusters like the Pleiades. The higher spot coverage may reflect either additional contributors to the lithium spread or systematic underestimates of spot coverage. Lithium abundance analysis in young clusters should therefore consider higher and distributed spot fractions rather than a single value.

Starspots can significantly affect the stellar age measurement. The lithium depletion boundary age may be erroneously underestimated by 15--30\% through radius inflation caused by starspots \citep{jackson2014effect,somers2015rotation}. Spots also shift the positions of stars on the HR diagram, which leads to an underestimation of the HR-diagram age \citep[e.g.][]{Jeffries2017vel,gully2017placing}. The discrepancy between the derived lithium depletion boundary age and HR diagram age can be largely reconciled by considering the influence of starspots \citep{Jeffries2017vel,binks2021ldb,binks2022ngc}, although correcting for its effects is still a challenge. Although some studies \citep[e.g.][]{jackson2014effect,Jeffries2017vel,somers2020spots,Franciosini2022pmsLi} consider the effects of spots, we still lack a complete assessment of spots versus mass and age.
Even in spectra, measurements are plagued by a degeneracy of the spot fraction and the spot temperature in an oversimplified two-component model. The problem of age measurement in young stars further impacts our understanding of stellar evolution and the timescale of disk evolution and planet formation. Future studies may combine spectra and multi-epoch photometry to solve the spot problem.

\subsection{{Predictions for the Beryllium Dispersion}}

{ 
In addition to $^7$Li, $^9$Be provides a complementary diagnostic of the role of starspots. $^9$Be is the only stable isotope of Beryllium, which is synthesized predominantly by cosmic ray spallation \citep{Reeves1970}. Because Galactic Be enrichment occurs on timescales much longer than the few-Myr formation interval of an open cluster, cluster member stars are expected to have nearly the same natal Be abundance. In stellar interiors, $^9$Be  is  destroyed via the proton capture  (p, $\alpha$) and (p, d) reactions \citep{Rapisarda2021}. Since the temperature requirement of these reactions is higher than the $^7$Li burning reaction $^7$Li(p, $^4$He)$^4$He \citep{Boesgaard2002Be}, stellar photospheric Be destruction requires a hotter layer thus a deeper mixing than Li destruction \citep{Xiong2007Be}. Such Be destruction mixing corresponding to stars at least $\sim$  100 Myr at Solar metallicity \citep{Montalban2000}. At the young age of NGC~2264, significant Be depletion is not expected under standard pre-main sequence models for its stellar mass range.}

{The only measurable transitions of $^9$Be is the \ion{Be}{2} resonance doublet around 3130\,\AA \citep{Wallerstein1969LiBe}. The near-UV continuum at these wavelengths is expected to be dominated by the warmer photospheric component, making the \ion{Be}{2} lines less sensitive to cool spots than the \ion{Li}{1} 6708\,\AA\ line.
Therefore, if the Li dispersion in NGC~2264 is primarily an atmospheric effect caused by spot-induced temperature inhomogeneities, stars of similar mass and age should exhibit a much smaller dispersion in their inferred Be abundances.}


{ 
The only comprehensive study of Be in young regions shows no difference in Be abundance in  IC~2391 and IC~2602 despite a substantial Li abundance spread \citep{Smiljanic2011LiBe}.  For older regions, some Be depletion is found in Pleiades \citep{Boesgaard2003Be} and Hyades \citep{Boesgaard2002Be}, but the dispersion of Be abundance is comparable to the measurement error and much smaller than the Li dispersion.}

{ 
The spread in \ion{Li}{1} equivalent widths and lack of spread in \ion{Be}{2} equivalent widths have been interpreted in the context of internal mixing.  However, this spread is also consistent with our scenario in which spots produce an apparent spread in Li equivalent width without a corresponding dispersion in Be. In principle, it would also be consistent with an early burning of Li, though this is not expected from models and no age spread has been identified here.  The difference between Li and Be may be inconsistent with some of the more exotic scenarios of the depletion or enhancement of Li abundance.}


\section{Conclusions}
{ 
In this paper, we have revisited the lithium-starspots connection and demonstrated that the spread in EW(Li) in NGC~2264 can be partially but not entirely explained by surface temperature inhomogeneities caused by starspots, using a WTTS sample from GES DR~5.1. The Li spread predicted by a simple empirical model can cover 68\% of the stars. The remaining dispersion suggests that additional factors may contribute, though we find no empirical correlation between age, gravity and Li dispersion in NGC~2264.
}

Using high-resolution CFHT spectra and ASAS-SN photometry, we investigate the correlation between EW(Li), TiO index and V-band variability for 15 WTTSs, where a clear correlation is found to support the interpretation that surface spot coverage modulates the apparent lithium line strength. This implies that the lithium spread in very young clusters can mainly result from differences in spot coverage among stars with similar effective temperatures.

{ 
We then construct an empirical starspot model by fitting the lower envelope of the EW(Li)--$T_{\rm phot}$ relation for WTTSs in NGC~2264. This lower edge is assumed to be unspotted.  We investigate the effect of starspots by adding a portion of EW(Li) from a cooler temperature to the observed EW(Li). The results show that the starspot model can reproduce much of the observed EW(Li) spread, and that larger temperature contrasts between the photosphere and the spots produce a larger lithium dispersion. 
}

{ 
We also investigate whether surface gravity and stellar age can contribute to the Li dispersion. We apply Monte Carlo simulation to estimate the Li dispersion, using the lithium depletion predicted by the SPOTS model \citep{somers2020spots} and the conversion from Li abundance to EW(Li) by the curves of growth \citep{Franciosini2022cog}. Model predicts that differences in surface gravity and stellar age among cluster members can produce certain Li spread, but we find no corresponding empirical relation between EW(Li) residuals and either surface gravity or inferred age.}

We conclude that starspots can be treated as a major contributor to the observed EW(Li) spread in NGC~2264, but do not rule out additional contributions (isochronal age spread, difference in surface gravity, or other activity-related effects). Moreover, the physical mechanism behind the lithium-rotation connection still cannot be directly examined by starspots. Further work is needed to better clarify the underlying physical mechanisms.

\section{Acknowledgements}

We thank Michael Gully-Santiago, Garrett Somers, Bharat Kumar, and Qiliang Fang for contributions in the early development of this paper. Yan thanks Yuxin Guo, Xiaoyi Ma, and Chenghui Zhao for their help and support during the course of this work. This manuscript benefited from the use of large language models (GPT, Gemini, and Claude) for code debugging, as well as language and grammar polishing.

YL and GJH are supported by National Natural Science Foundation of China (NSFC) general program 12573031. 
{ 
This work made use of data obtained from the ESO Science Archive Facility (DOI: \url{https://doi.org/10.18727/archive/25}).
}
XF acknowledges the support of the NSFC No. 12573040, No. 12533008, and the China Manned Space Program with grant No. CMS-CSST-2025-A14.

\clearpage

\appendix
\section{Selecting WTTSs in NGC~2264}
\label{app:A}

\begin{figure*}[h]
    \centering
    \includegraphics[width=0.81\textwidth]{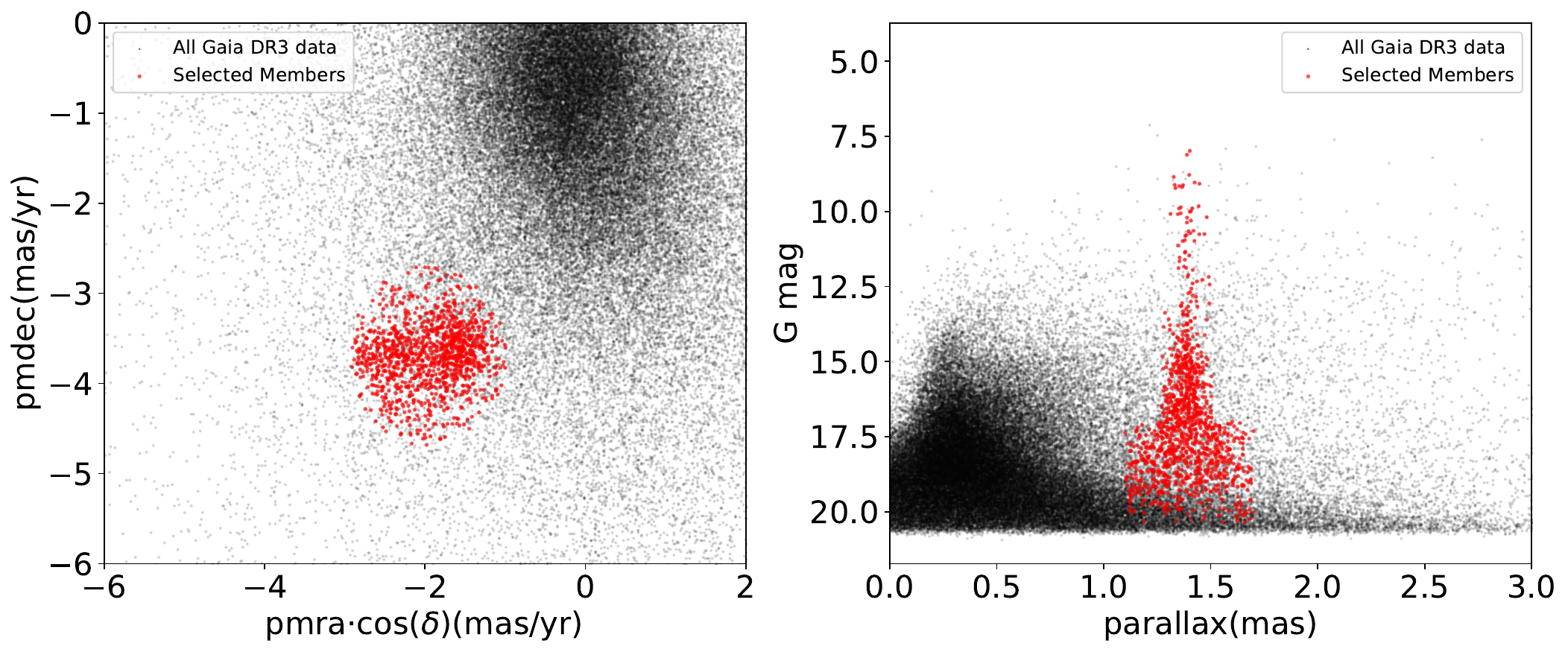}
    \includegraphics[width=0.8\textwidth]{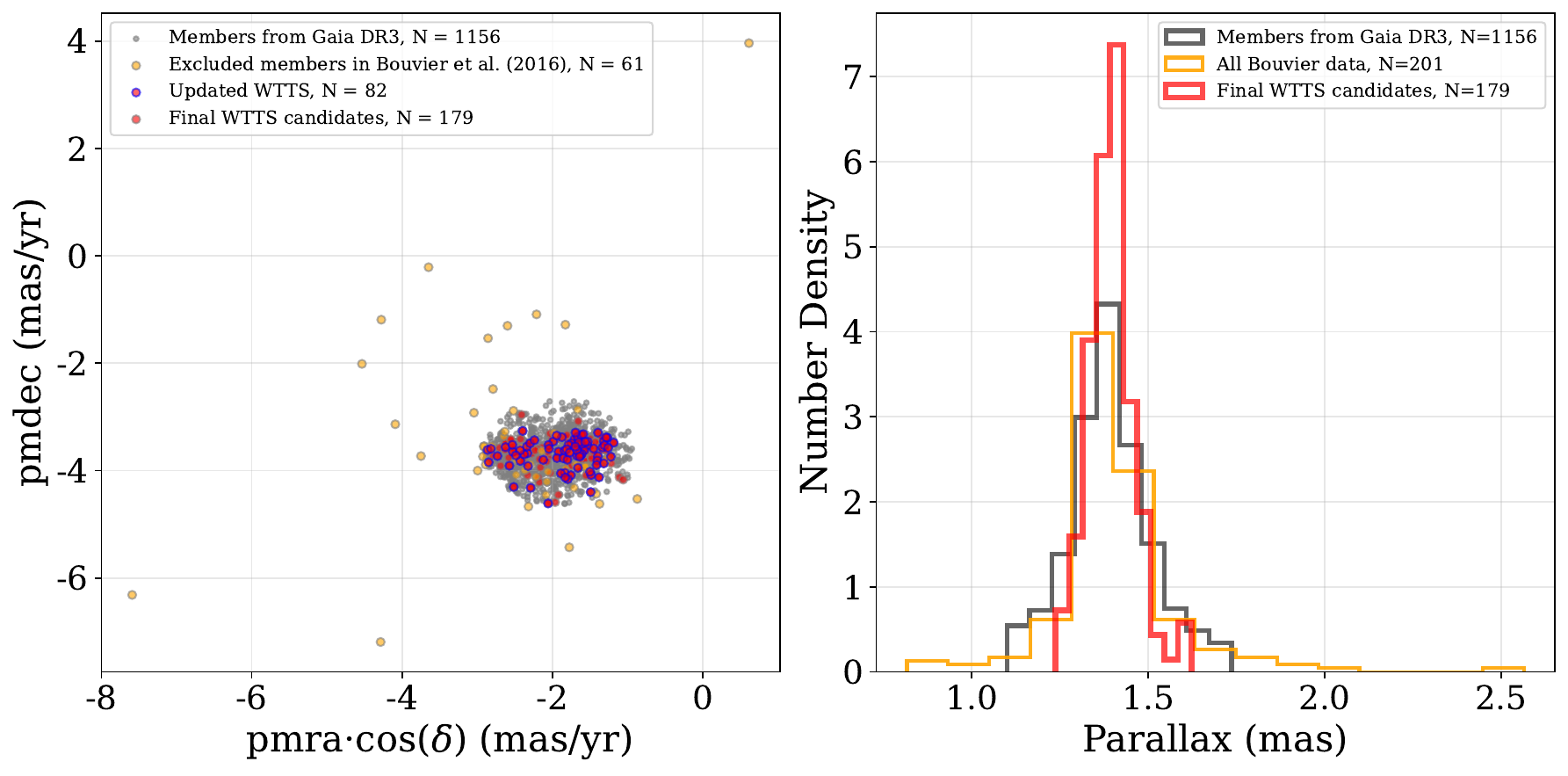}
    \caption{NGC~2264 member selection from Gaia DR3. The upper panels show the proper-motion distribution and the $G$-band magnitude versus parallax diagram. The black points denote all Gaia DR3 sources within $1^\circ$ of the cluster center, and the red points show the selected NGC~2264 members. The lower panels compare the proper-motion and parallax distributions.}
    \label{fig:pm_parallax_NGC2264}
\end{figure*}

To obtain a high confidence sample of NGC~2264 members from Gaia~DR3, we apply a series of proper-motion and parallax selections. Our priority is to optimize membership probability, which will exclude some real members.
Our procedure follows the criteria adopted by \citet{bu2025disk}, which are summarized below.

We first retrieve all Gaia~DR3 sources within $1^\circ$ of the center of NGC~2264. We then select stars with photometric uncertainties $\sigma_G < 0.05$ mag, proper-motion uncertainties $\sigma_{\rm pmRA},\,\sigma_{\rm pmDec}<1$ mas\,yr$^{-1}$, and positive parallaxes. Candidate members are required to satisfy

\[
\sqrt{({\rm pmRA}\cos\delta+2.0)^2+({\rm pmDec}+3.7)^2}<1.0
\ {\rm mas\,yr^{-1}},
\]

and

\[
1.1 < \varpi < 1.8 \ {\rm mas},
\]

corresponding to a central distance of approximately 720 pc and a distance range of 560--910 pc. These criteria yield 1239 candidate members.

To further refine the sample, we perform iterative sigma-clipping in parallax space. At each iteration, we compute the median parallax, $\tilde{\varpi}$, and its standard deviation, $\sigma_\varpi$. We retain stars satisfying

\[
|\varpi-\tilde{\varpi}| < 1\sigma_\varpi
\]

for $G\le17$, and

\[
|\varpi-\tilde{\varpi}| < 3\sigma_\varpi
\]

for $G>17$, since Gaia parallaxes become significantly less precise at fainter magnitudes \citep{brown2021gaia}. After convergence, 1156 stars remain and are adopted as the final Gaia-selected members of NGC~2264. Figure~\ref{fig:pm_parallax_NGC2264} shows that the cluster exhibits two proper-motion populations. However, these two kinematic substructures appear to have negligible age differences \citep{Cheshire2025NS2264} and therefore do not bias our lithium-spread analysis.

{ 
To construct the final sample for this work, we cross-match the updated GES DR5.1 catalogue with our Gaia-selected members and apply additional criteria to identify bona fide WTTSs. 415 stars have EW(Li) and temperature measurements.}

{ 
We require radial velocities from GES DR~5.1 within 30\,km\,s$^{-1}$ to ensure kinematic membership. We then remove disk-bearing stars using infrared excess in 2MASS $J$ and WISE $W1/W2$ photometry. For each of the two color-color diagrams, $J-W1$ and $J-W2$ versus BP$-$RP, we fit a cubic spline to the photospheric locus and identify stars lying more than $3\sigma$ above the locus as disk candidates. Stars with excess $J-W1$ or $J-W2$ are excluded. We also exclude stars without $J-W1$ or $J-W2$  measurement to ensure a strict WTTS sample. 
}

{ 
We then remove accreting stars using two H$\alpha$ diagnostics: the spectral type-dependent EW(H$\alpha$) criterion and the 10\% H$\alpha$ width > 270\,km\,s$^{-1}$ criterion \citep{White2003ha}. We further remove stars previously identified as Class~II objects in \citet{Venuti2018av}. Stars that do not satisfy either criterion are excluded.}

{ 
The final 179 high-confidence WTTS sample therefore consists of stars with measured EW(Li) and spectroscopic temperatures, no infrared excess in either $J-W1$ or $J-W2$, no possible H$\alpha$ accretion signatures, no Class~II classification, and radial velocities consistent with cluster membership. 
}

{ 
Of the 201 Bouvier et al. (2016) stars, 188 of them are in the GES DR5.1 catalog with EW(Li) and temperatures
measurements, 50 are excluded as lower confidence members based on Gaia DR3 astrometry, 24 are reclassified as 
non-WTTS (20 showing H$\alpha$ accretion, 4 with infrared excess, and 5 classified as Class~II in \citealt{Venuti2018av}), and 17 lack the required infrared photometry, leaving 97 stars in common with our final sample. The remaining 82 WTTSs are newly identified from GES DR5.1. We emphasize that these exclusions reflect our strict selection criteria rather than problems with the original classification; the final sample is conservative and incomplete, but provides a clean WTTS census for studying the lithium spread.
Figure~\ref{fig:pm_parallax_NGC2264} compares the proper-motion and parallax distributions of the final sample, the WTTS sample of \citet{bouvier2016gaia}, and the rejected sources. The Li dispersion does not change much after updating the sample. Updating the sample has little effect on the overall shape of the lithium dispersion. We find no systematic differences in the EW(Li) distribution as a function of either proper motion or spatial position.
}

\newpage
\section{The Lithium--Starspots Connection}
\begin{table*}[!h]
    \centering
    \footnotesize
    \caption{{Spectral types, rotation periods, peak-to-peak variability amplitudes for EW(Li), TiO index, and V-band magnitude, and Spearman's rank correlation correlation coefficients with $p$-values among these quantities for the 15 WTTS sample. }}
    \label{tab:wtts_correlation}
    \begin{tabular}{lcccccccc}
    \toprule
    Source Name & SpT & P(d) & $\rm EW_{Li}$ Var.\ (m\AA) & TiO Var. & Vmag Var. & $\rho_{\rm EW-Vmag}\ (p)$ & $\rho_{\rm TiO-Vmag}\ (p)$ & $\rho_{\rm EW-TiO}\ (p)$ \\
    \midrule
    V410 Tau & K3.0 & 1.872 & 13$\pm$2 & 0.007$\pm$0.001 & 0.074$\pm$0.013 & +0.33\ (0.007) & +0.43\ ($< 0.001$) & +0.21\ (0.098) \\
    Par 1379 & K4.0 & 5.676 & 10$\pm$3 & 0.006$\pm$0.002 & 0.063$\pm$0.007 & +0.46\ (0.003) & +0.36\ (0.023) & +0.01\ (0.957) \\
    TaP 26 & K4.0 & 0.714 & 14$\pm$2 & 0.013$\pm$0.000 & 0.103$\pm$0.008 & +0.29\ ($< 0.001$) & +0.65\ ($< 0.001$) & +0.24\ ($< 0.001$) \\
    ROX 39 & K5.0 & 0.882 & 8$\pm$4 & 0.007$\pm$0.001 & 0.077$\pm$0.008 & +0.01\ (0.913) & +0.26\ (0.010) & +0.11\ (0.307) \\
    ROXs 45F & K5.0 & 2.454 & 14$\pm$3 & 0.032$\pm$0.002 & 0.060$\pm$0.008 & +0.59\ ($< 0.001$) & +0.82\ ($< 0.001$) & +0.52\ ($< 0.001$) \\
    TaP 45 & K6.0 & 9.587 & $-7\pm$4 & 0.026$\pm$0.003 & 0.089$\pm$0.009 & -0.22\ (0.161) & +0.64\ ($< 0.001$) & -0.23\ (0.146) \\
    V830 Tau & K7.5 & 2.744 & 14$\pm$1 & 0.032$\pm$0.001 & 0.188$\pm$0.008 & +0.53\ ($< 0.001$) & +0.72\ ($< 0.001$) & +0.61\ ($< 0.001$) \\
    V819 Tau & K8.0 & 5.535 & 24$\pm$2 & 0.075$\pm$0.002 & 0.415$\pm$0.007 & +0.89\ ($< 0.001$) & +0.96\ ($< 0.001$) & +0.87\ ($< 0.001$) \\
    RX J1608.0-3857 & M0.0 & 2.425 & 11$\pm$2 & 0.046$\pm$0.001 & 0.124$\pm$0.008 & +0.53\ ($< 0.001$) & +0.78\ ($< 0.001$) & +0.49\ ($< 0.001$) \\
    TWA 6 & M0.0 & 0.541 & 44$\pm$3 & 0.073$\pm$0.001 & 0.275$\pm$0.007 & +0.63\ ($< 0.001$) & +0.92\ ($< 0.001$) & +0.58\ ($< 0.001$) \\
    RX J1609.5-3850 & M0.5 & 3.868 & 29$\pm$3 & 0.093$\pm$0.002 & 0.172$\pm$0.008 & +0.73\ ($< 0.001$) & +0.83\ ($< 0.001$) & +0.79\ ($< 0.001$) \\
    TWA 25 & M0.5 & 5.065 & 30$\pm$2 & 0.130$\pm$0.002 & 0.407$\pm$0.008 & +0.76\ ($< 0.001$) & +0.84\ ($< 0.001$) & +0.93\ ($< 0.001$) \\
    LkCa 7 & M1.2 & 5.647 & 33$\pm$2 & 0.197$\pm$0.002 & 0.427$\pm$0.007 & +0.87\ ($< 0.001$) & +0.89\ ($< 0.001$) & +0.75\ ($< 0.001$) \\
    LkCa 4 & M1.3 & 3.374 & 20$\pm$4 & 0.122$\pm$0.002 & 0.309$\pm$0.007 & +0.58\ ($< 0.001$) & +0.92\ ($< 0.001$) & +0.63\ ($< 0.001$) \\
    TWA 8a & M2.9 & 4.630 & 4$\pm$4 & 0.071$\pm$0.008 & 0.098$\pm$0.008 & +0.21\ (0.107) & +0.72\ ($< 0.001$) & +0.40\ (0.002) \\
    \bottomrule
    \end{tabular}
\end{table*}

\newpage

\figsetstart
\figsetnum{11}
\figsettitle{Lithium--Starspot Relation for the WTTS Sample}

\figsetgrpstart
\figsetgrpnum{11.1}
\figsetgrptitle{LkCa 7}
\figsetplot{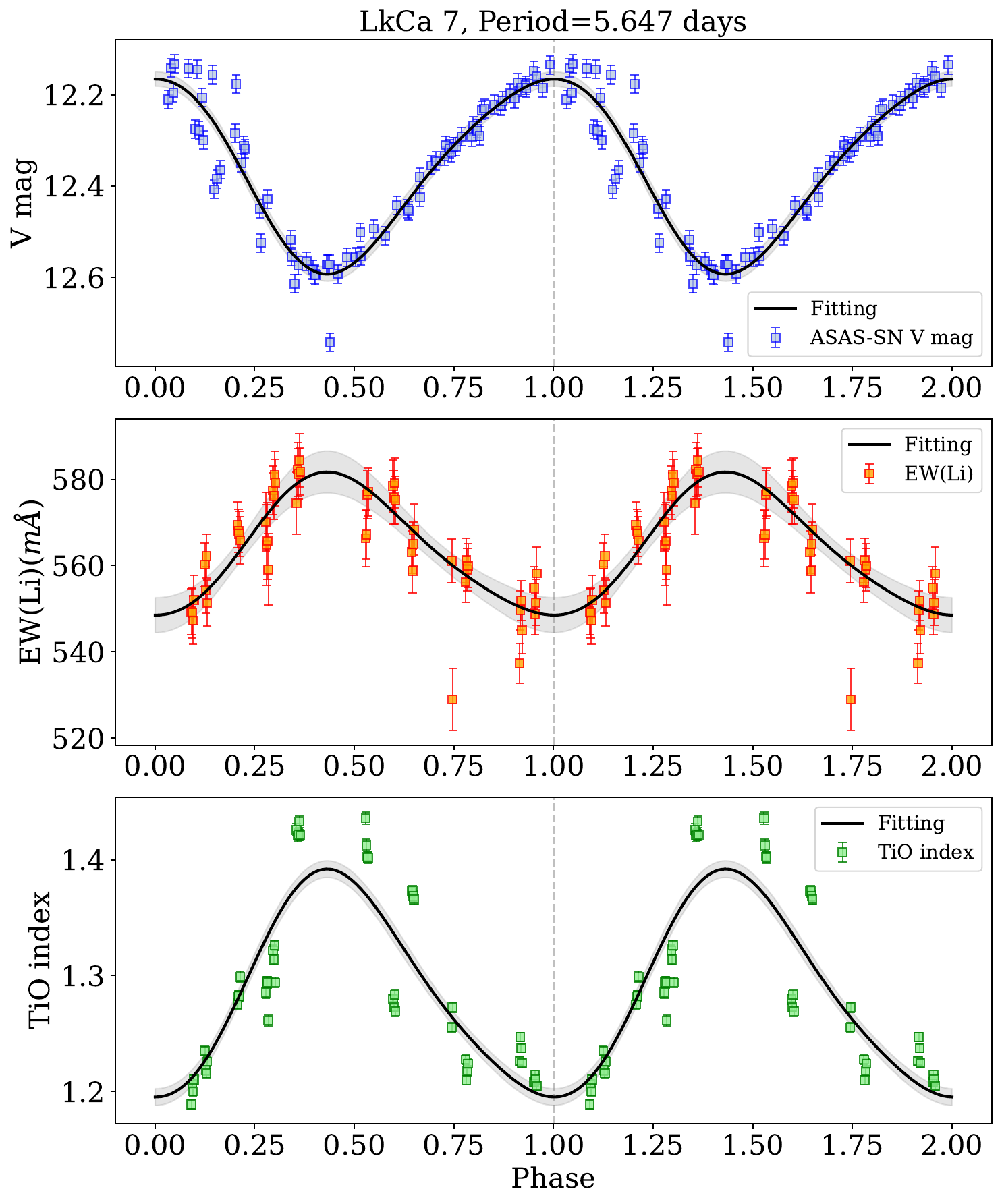}
\figsetplot{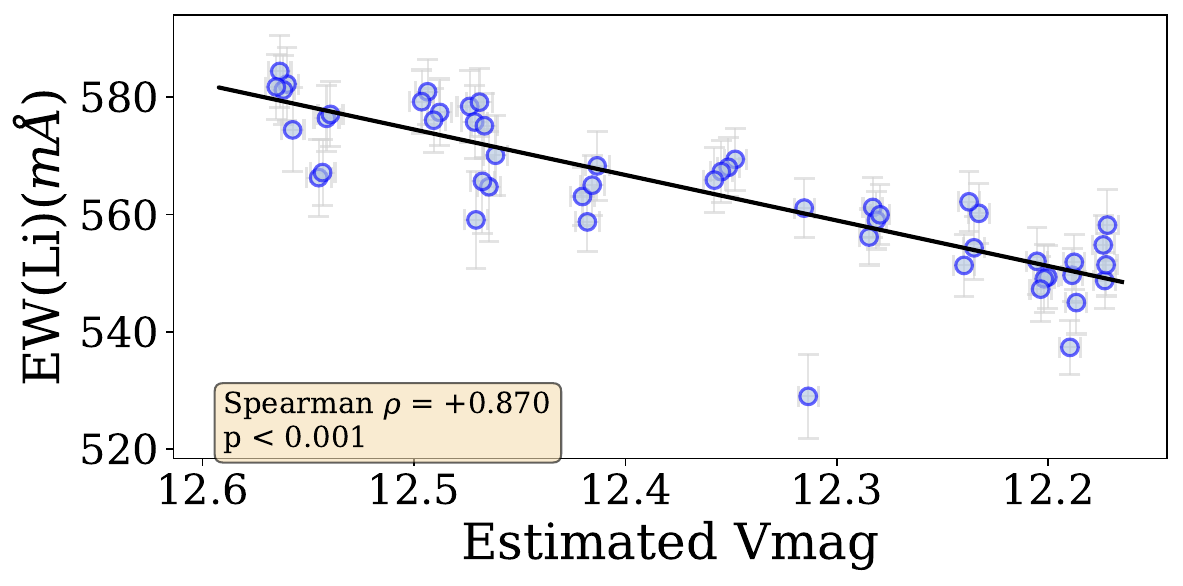}
\figsetplot{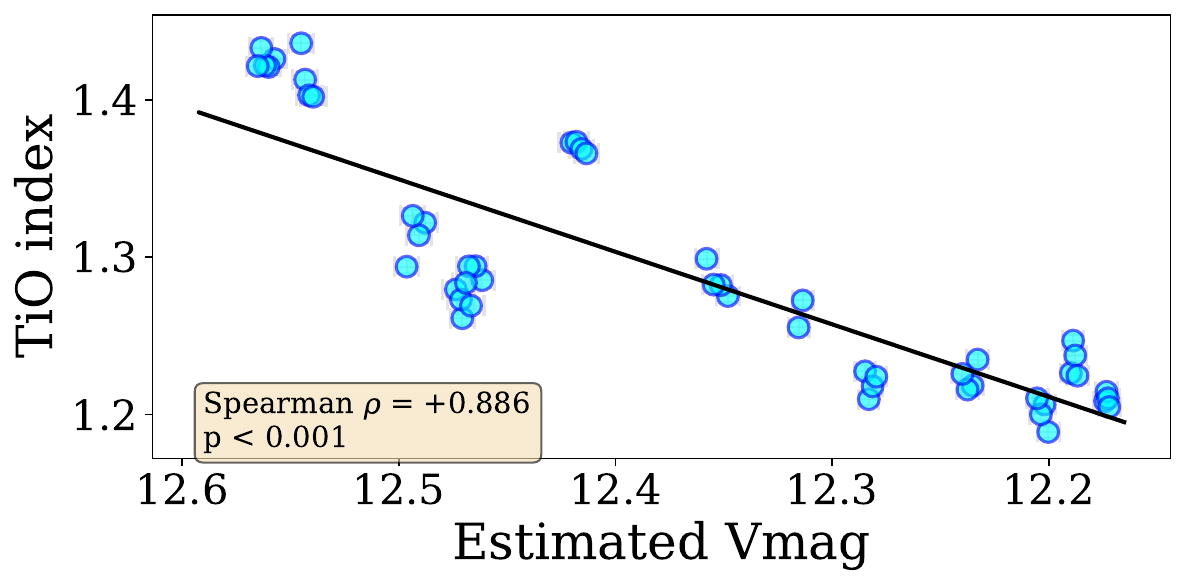}
\figsetplot{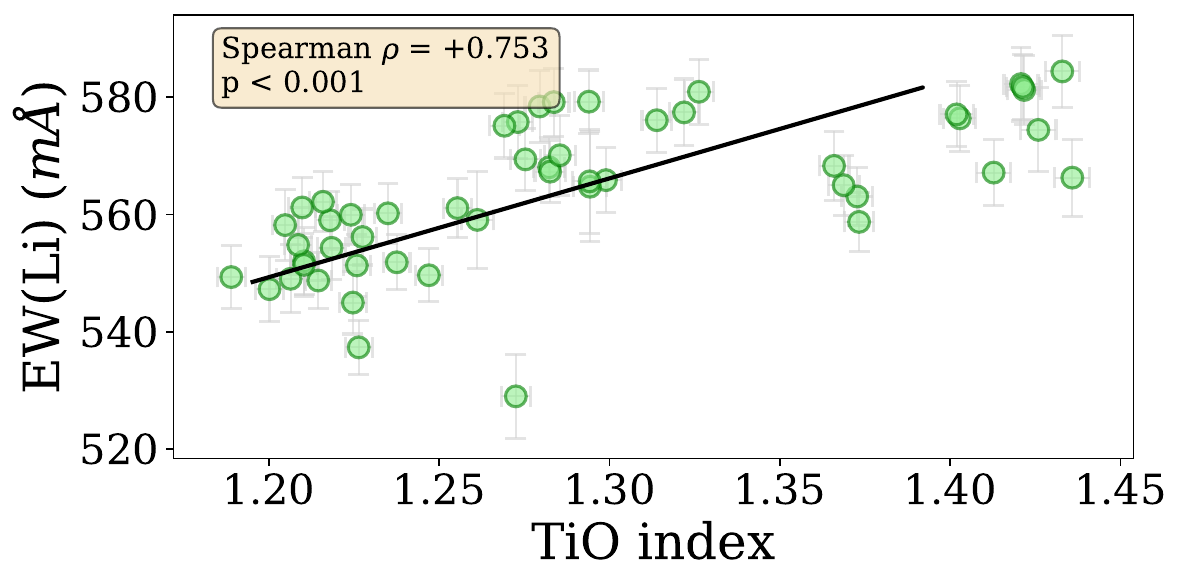}
\figsetgrpnote{Lithium--starspot diagnostics for LkCa 7. Left: variation of $V$-band magnitude, EW(Li), and TiO index with phase. Top right: EW(Li) vs.\ $V$-band magnitude. Middle right: TiO index vs.\ $V$-band magnitude. Bottom right: change in EW(Li) and TiO index with phase.}
\figsetgrpend

\figsetend

\begin{figure*}[h]
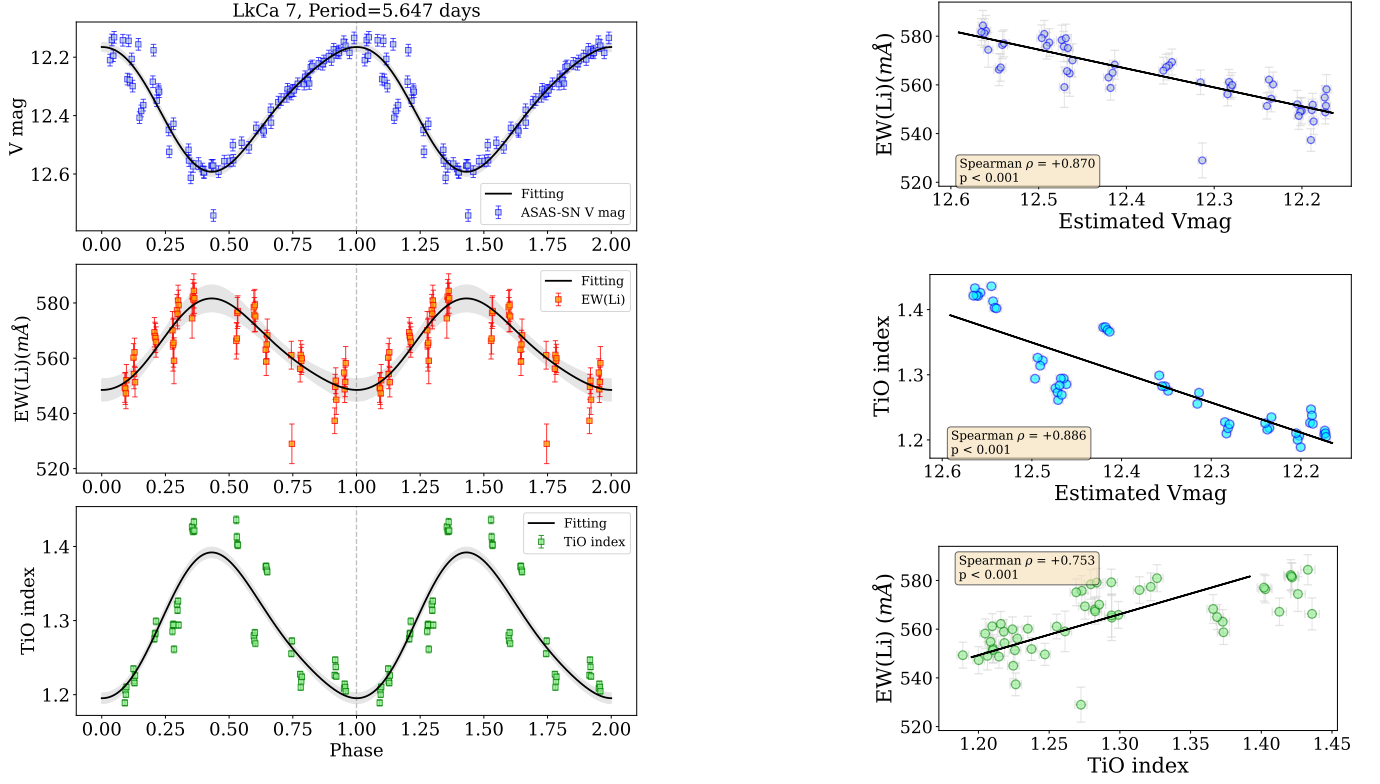

    \centering
    \subfigure{
    \begin{minipage}{0.47\textwidth}
        \includegraphics[width=\textwidth]{variability_compare_lkca7.pdf}
        \label{fig:variability_compare_lkca7}
    \end{minipage}
    }
    \hfill
    \begin{minipage}{0.36\textwidth}
    \subfigure{
        \includegraphics[width=\textwidth]{correlation_test_lkca7.pdf}
        \label{fig:correlation_test_lkca7}
    }
    \subfigure{
        \includegraphics[width=\textwidth]{TiO_correlation_lkca7.pdf}
        \label{fig:TiO_correlation_lkca7}
    }
    \subfigure{
        \includegraphics[width=\textwidth]{Li_vs_TiO_lkca7.pdf}
        \label{fig:Li_vs_TiO_lkca7}
    }
    \end{minipage}
    \caption{Lithium--starspot diagnostics for LkCa~7, used here as the representative example. 
    Left: variation of $V$-band magnitude, EW(Li), and TiO index with phase. 
    Top right: EW(Li) vs.\ $V$-band magnitude. Middle right: TiO index vs.\ $V$-band magnitude. 
    Bottom right: change in EW(Li) and TiO index with phase. The complete figure set (14 images) is available in the online journal.}
    \label{fig:li_spot_lkca7}
\end{figure*}

\clearpage
\section{Different Temperature Contrast of the Starspots Model}
\label{APP:C}
Section 4.2 describes how the EW(Li) depends on spot parameters, with fixed spot temperature relative to the photosphere, $\log T_{\rm phot} - \log T_{\rm spot} = 0.12$.  These plots show the same curves with other methods to assess spot temperature.

\begin{figure*}[h]
    \centering
    \subfigure{
    \begin{minipage}{0.48\textwidth}
    \subfigure{
        \includegraphics[width=\textwidth]{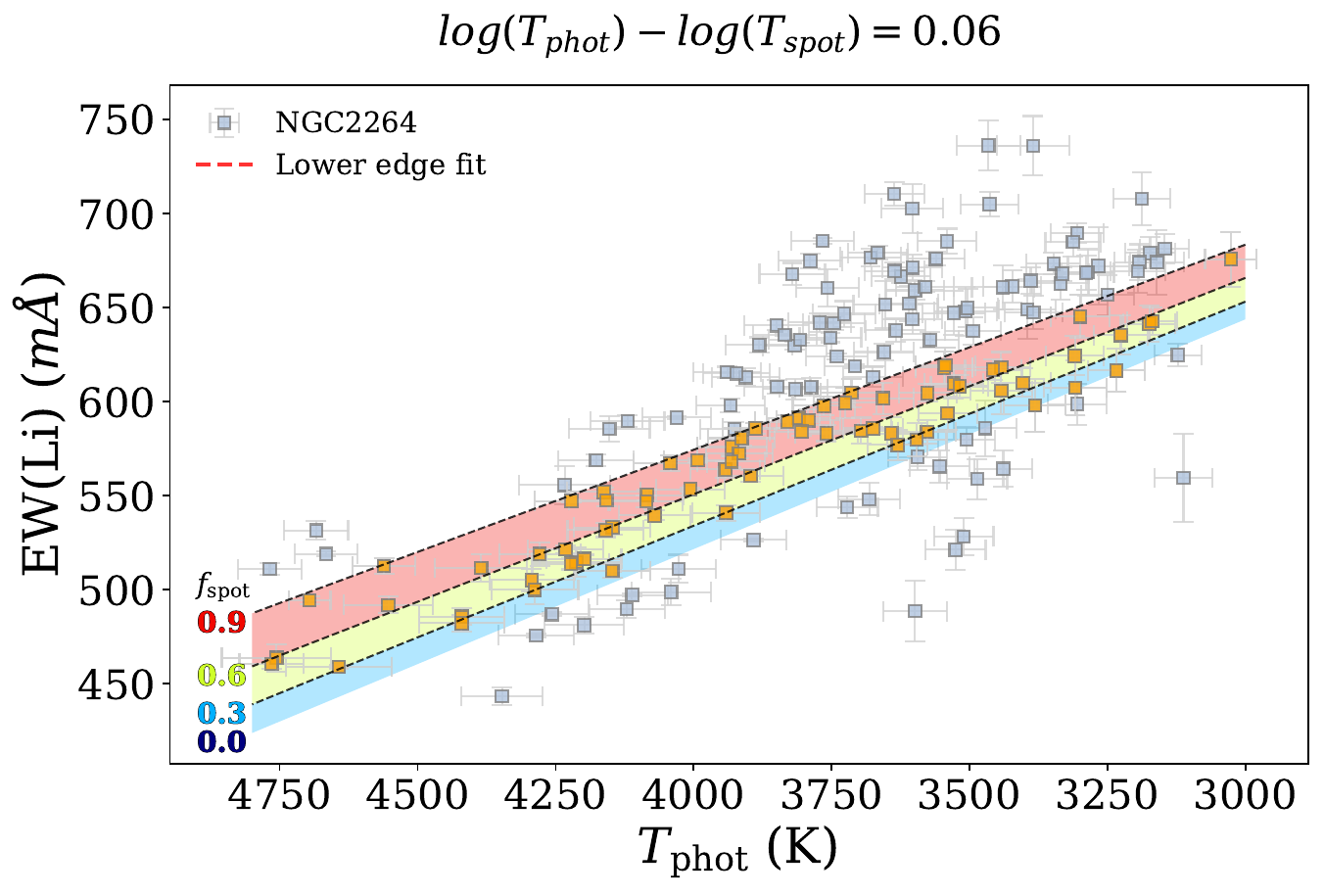}
    }
    \subfigure{
        \includegraphics[width=\textwidth]{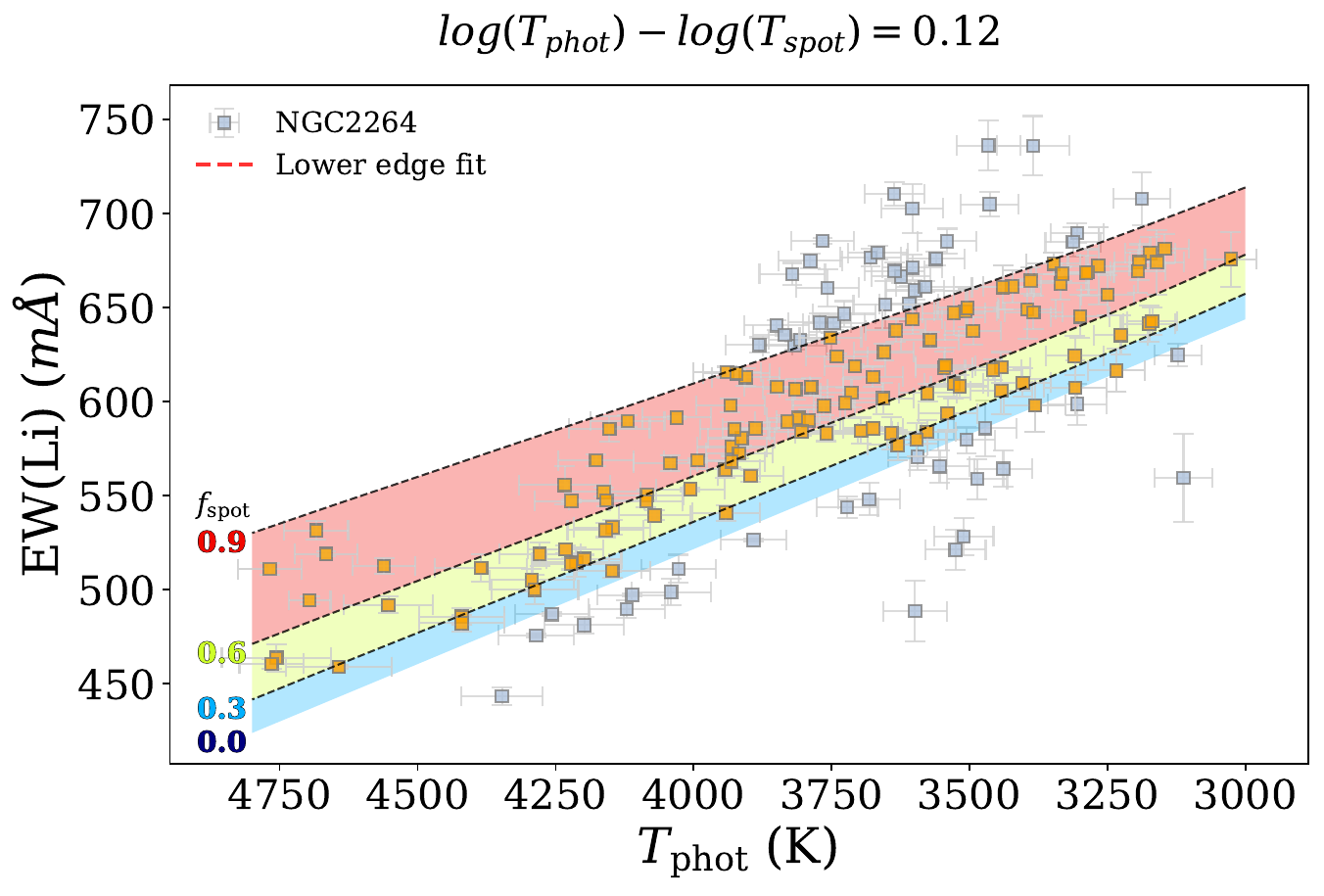}
    }
    \subfigure{
        \includegraphics[width=\textwidth]{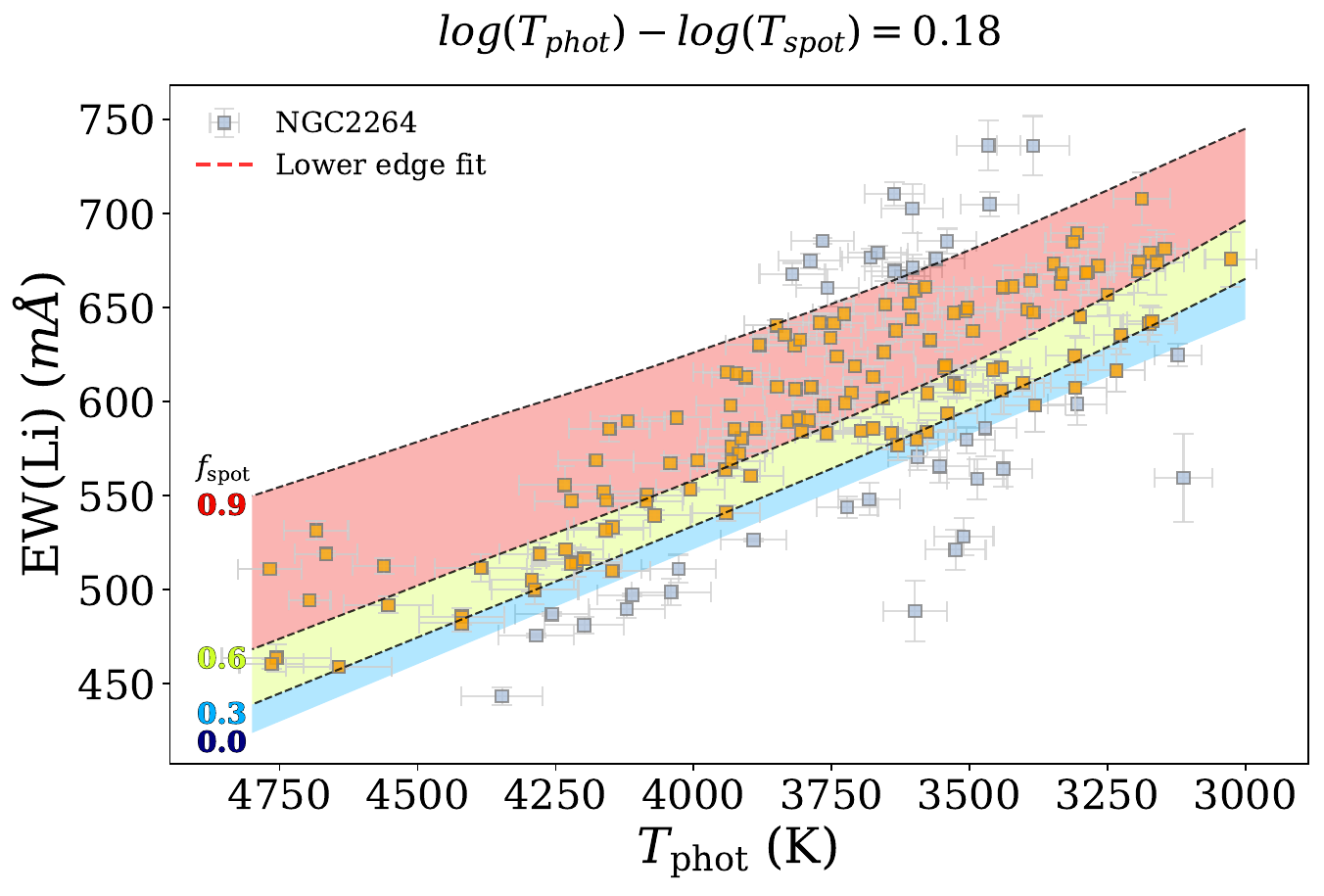}
    }
    \end{minipage}
    }
    \hfill
    \begin{minipage}{0.48\textwidth}
    \subfigure{
        \includegraphics[width=\textwidth]{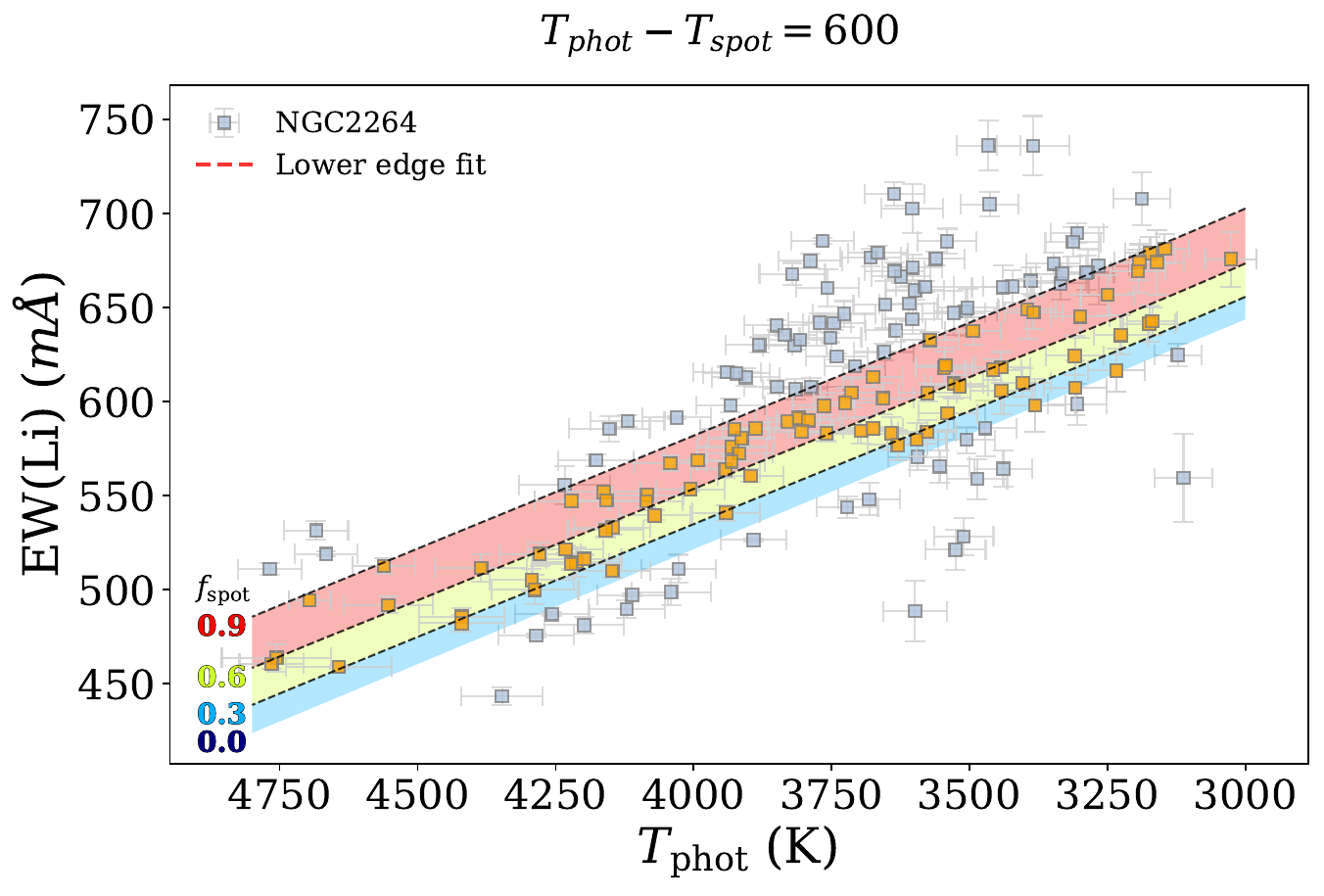}
    }
    \subfigure{
        \includegraphics[width=\textwidth]{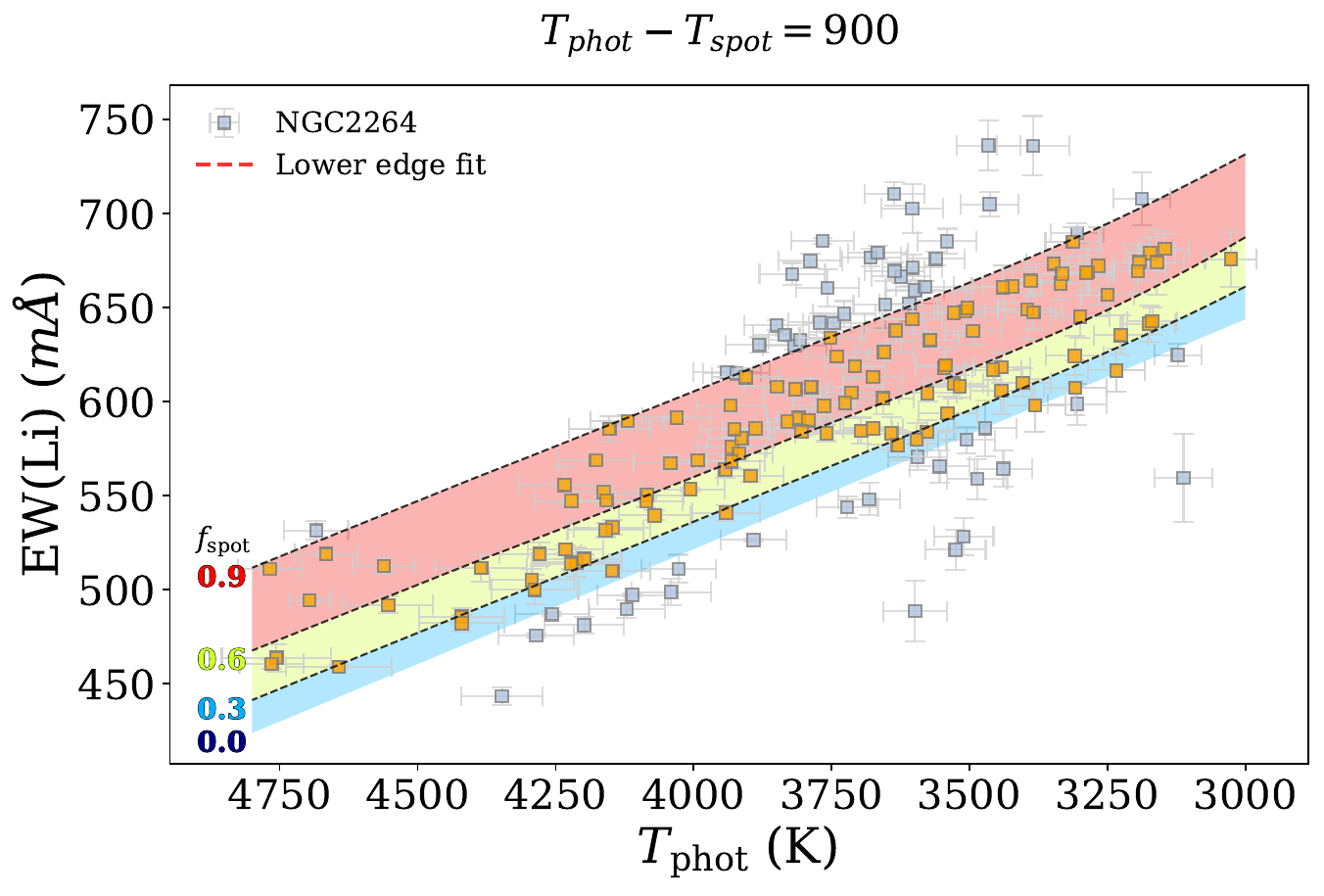}
    }
    \subfigure{
        \includegraphics[width=\textwidth]{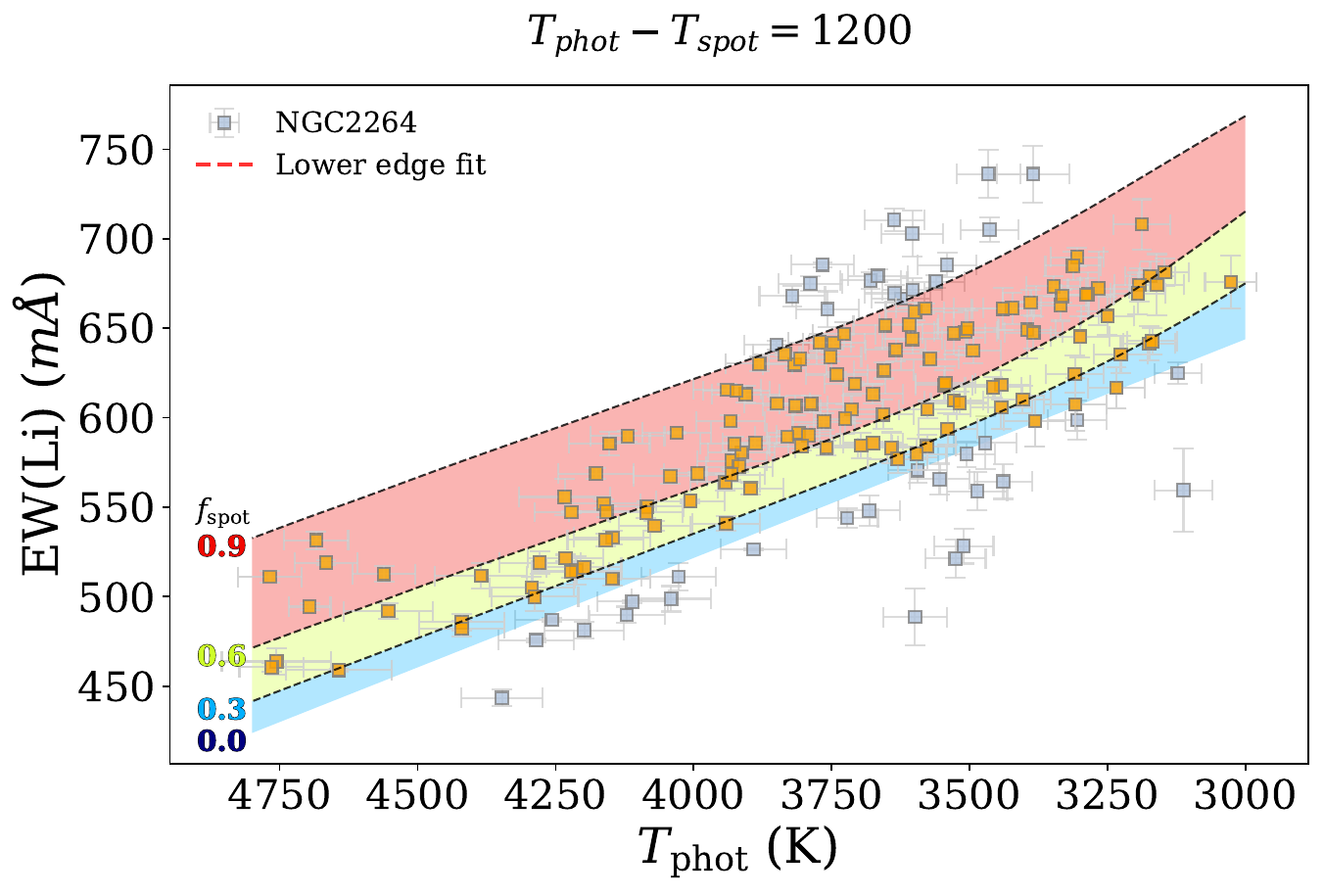}
        }
    \end{minipage}
    \label{fig:li_spot_model_Tspot_dex}
    \caption{Different temperature contrasts influence on the starspot model. Left: $T_{\rm{phot}} / T_{\rm{spot}}= \rm{const}$; Right: $T_{\rm{phot}}-T_{\rm{spot}}=\rm{const}$. From top to bottom, the temperature contrast increases, which leads to wider lithium spreads. In this paper, We apply $\log(T_{\rm spot})=\log(T_{\rm phot})-0.12$ in the starspots model.}
    
\end{figure*} 


\clearpage

\bibliographystyle{apj}
\bibliography{lib}

\end{CJK*}
\end{document}